\documentclass[fleqn,usenatbib]{mnras}

\usepackage[T1]{fontenc}

\DeclareRobustCommand{\VAN}[3]{#2}
\let\VANthebibliography\thebibliography
\def\thebibliography{\DeclareRobustCommand{\VAN}[3]{##3}\VANthebibliography}

\usepackage{graphicx}	
\usepackage{amsmath}	
\usepackage{natbib}
\usepackage{amssymb}
\usepackage{enumitem}
\usepackage{amsmath}
\usepackage{iondefs}

\usepackage{newtxtext,newtxmath}

\title[Examining quasar absorption-line analysis methods]{Examining quasar absorption-line analysis methods: the tension between simulations and observational assumptions key to modelling clouds}

\author[\sc Marra {\etal}]
{Rachel Marra,$^{1}$\thanks{E-mail: rmarra@nmsu.edu}
Christopher W. Churchill,$^{1}$ 
Glenn G. Kacprzak,$^{2,3}$
Nikole M. Nielsen,$^{2,3}$
Sebastian \newauthor Trujillo-Gomez, $^{4}$
and James G. Lewis $^{5}$
\\
$^{1}$Department of Astronomy, New Mexico State University, Las Cruces, NM 88003, USA\\
$^{2}$Centre for Astrophysics and Supercomputing, Swinburne University of Technology, Hawthorn, Victoria 3122, Australia\\
$^{3}$ARC Centre of Excellence for All Sky Astrophysics in 3  Dimensions (ASTRO 3D), Australia\\
$^{4}$ Astronomisches Rechen-Institut, Zentrum f{\"u}r Astronomie der Universit{\"a}t Heidelberg, Monchhofstra{\ss}e 12-14, D-69120 Heidelberg, Germany\\
$^{5}$ Department of Mathematics, Cornell University, Ithaca, NY 14850, USA\\
}

\date{Accepted XXX. Received YYY; in original form ZZZ}

\pubyear{2021}

\begin{document}
\label{firstpage}
\pagerange{\pageref{firstpage}--\pageref{lastpage}}
\maketitle

\begin{abstract}
A key assumption in quasar absorption line studies of the circumgalactic medium (CGM) is that each absorption component maps to a spatially isolated ``cloud'' structure that has single valued properties (e.g. density, temperature, metallicity). We aim to assess and quantify the degree of accuracy underlying this assumption. We used adaptive mesh refinement hydrodynamic cosmological simulations of two $z=1$ dwarf galaxies and generated synthetic quasar absorption-line spectra of their CGM. For the {\SiII}~$\lambda 1260$ transition, and the {\CIVdblt} and {\OVIdblt} fine-structure doublets, we objectively determined which gas cells along a line-of-sight (LOS) contribute to detected absorption. We implemented a fast, efficient, and objective method to define individual absorption components in each absorption profile. For each absorption component, we quantified the spatial distribution of the absorbing gas. We studied a total of 1,302 absorption systems containing a total of 7,755 absorption components. 48\% of {\SiII}, 68\% of {\CIV}, and 72\% of {\OVI} absorption components arise from two or more spatially isolated ``cloud'' structures along the LOS. Spatially isolated ``cloud'' structures were most likely to have cloud-cloud LOS separations of 0.03$R_{vir}$, 0.11$R_{vir}$, and 0.13$R_{vir}$ for {\SiII}, {\CIV}, and {\OVI}, respectively. There can be very little overlap between multi-phase gas structures giving rise to absorption components. If our results reflect the underlying reality of how absorption lines record CGM gas, they place tension on current observational analysis methods as they suggest that component-by-component absorption line formation is more complex than is assumed and applied for chemical-ionisation modelling.
\end{abstract}


\section{Introduction}

Quasar absorption lines provide the vast majority of our real-world information for inferring how galaxies evolve through the accretion of intergalactic gas, expulsion of stellar driven winds, and feedback and recycling of gas that regulates the formation of their stars. An observational gateway to understanding this ``baryon cycle'' is the circumgalactic medium (CGM), the extended gaseous halo surrounding galaxies. The absorption lines from the CGM often reflect dynamically moving, clumpy gas structures, which manifest as complex multi-component absorption profiles. Key observational quantities include the kinematics, spatial distributions, metallicities, densities, and temperatures of the inflowing, outflowing, and recycling gas \citep[e.g.,][]{stocke13, werk14, lehner14, Lehner18, lehner19, Wotta16, Wotta19,  prochaska17, Pointon_2019}. These quantities are indirectly obtained from the absorption line data through chemical-ionisation models of the gas.  The modelling inputs include incomplete information (number of absorption components, column densities, etc.) measured from various absorbing transitions of various ions. Furthermore, as with all modelling, simplifying assumptions are required. For quasar absorption line studies, one of the key assumptions has been that each absorption component maps to a spatially isolated ``cloud'' structure that has single valued properties, such as density, temperature, and metallicity. 

From a theoretical standpoint, modern hydrodynamic cosmological simulations hold great potential for informing us how to interpret CGM gas observed with quasar absorption lines. Analysis of simulated CGM gas absorption properties, coupled with direct comparison of the simulated gas kinematics, spatial distribution, and physical conditions, is a powerful methodology for studying the baryon cycle. Such studies provide an open road for applying observational quasar absorption line analysis methods to simulated galaxies \citep[e.g.,][]{kacprzak10, churchill15, liang18, Kacprzak_2019, peeples19, peroux20, marra21a, marra21b, strawn21}. The trouble is, several insights gained from absorption lines studies of simulations also place tension on the analysis methods adopted by observers, i.e., the aforementioned assumptions that each absorption component maps to a spatially isolated iso-thermal cloud. Simulations show the absorbing gas structures are far more complex \citep{churchill15, liang18, peeples19, marra21b}. For example, \citet{churchill15} found that {\MgII} absorption and the bulk of {\HI} absorption arise in ``cloud-like'' structures that are confined within contiguous gas cells over small spatial scales. Though these structure have a narrow temperature distribution, they exhibit density and metallicity gradients as a function of line of sight (LOS) velocity.  On the other hand, {\CIV} and {\OVI} absorbing gas structures are not ``cloud like''; the absorption arises from gas spatially distributed over $\sim\! 100$~kpc with a complex velocity flow that results in multiple highly-separated locations giving rise to a single absorption component at a given value of LOS velocity. 

These insights present a significant challenge to analysis methods that employ Voigt profile (VP) modelling of the absorption profiles followed by chemical ionisation modelling constrained by the VP parameters, particularly when the analysis is based on the assumption that each VP component maps to a single discrete cloud having a single valued density, temperature, and metallicity.  

Some observational studies have explored chemical-ionisation modelling methods that treat the VP parameters of low- and high-ionisation absorption data (especially {\OVI} absorption) separately, thereby attempting to capture the multi-phase characteristics of CGM gas \citep[e.g.,][]{muzahid15, Zahedy19, Zahedy21, haislmaier21}. However, these methods force the ionisation modelling to conform to the VP fitted parameters, which may lead to systematic inaccuracies in the modelling outcomes (gas densities, temperatures, and metallicities) because the VP modelling is not informed by the gas physics but only by the kinematics of the absorption profiles. Methods that model the LOS velocity aligned component-by-component absorption of multiple ions while simultaneously performing the ionisation modelling can introduce an added flexibility in that they can treat each absorption component of a given ion as arising from a combination of multiple gas clouds, each with a unique density, temperature, and metallicity \citep[e.g.,][]{cwc_charlton, charlton03, ding05, Sameer21, Sameer22, nielsen22}. 

To increase our collective understanding of the underlying physical conditions of CGM gas that give rise to absorption lines in quasar spectra so that we can glean improved insight into chemical-ionisation modelling, we use mock spectra through hydrodynamic cosmological simulations of dwarf galaxies to  statistically characterise the spatial and mass distributions of low- and high-ionisation CGM gas observed in individual absorption components in complex absorption profiles. To this end, we locate thousands of mock LOS through the simulated galaxies and generate synthetic absorption line spectra for each LOS.  In this paper, we analyse the absorption profiles from the {\SiII}~$\lambda 1260$ transition, and the {\CIVdblt} and {\OVIdblt} fine-structure doublets. Absorption from these three ions allows us to explore and compare different phases of CGM gas.

In Section~\ref{Simulations}, we present the cosmological simulations. In Section~\ref{Methodology}, we describe the creation of the synthetic spectra from sight lines through the simulations and the analysis of the simulated {\SiII}, {\CIV}, and {\OVI} absorption profiles. In Section \ref{Results}, we present our spatial distribution of absorbing gas clouds. We discuss our findings in the context of quasar absorption line studies in Section \ref{Discussion}. We summarise our findings and provide concluding remarks in Section \ref{Conclusion}. Throughout we adopt an $H_{0} = 70$~{\kms}~Mpc$^{-1}$, $\Omega_{\tM} = 0.3$, $\Omega_{\Lambda} = 0.7$ cosmology.

\section{The Simulations}
\label{Simulations}

We studied two simulated low-mass dwarf galaxies at a redshift of $z = 1$ from the simulations of \citet{Trujillo15}. The basic properties of these galaxies are listed in Table~\ref{tab:galaxy}. The galaxies were simulated using the Hydrodynamic Adaptive Refinement Tree code known as hydro{\sc ART} \citep{Kravstov97, Kravstov99_thesis, Kravtsov_2003, ceverino09, Trujillo15}. The code combines an Eulerian treatment of hydrodynamics while employing the zoom-in technique of \citet{Klypin2001} to model a single galaxy at high resolution while capturing the large-scale cosmological environment. Details of the physical processes implemented in the simulations and the stellar formation and feedback recipes are given in \citet[][]{Trujillo15}. 

In brief, physical processes implemented in the code include star formation, supernovae feedback, Type II and Type Ia supernova metal enrichment, photoheating in {\HII} regions, radiation pressure, and metallicity-dependent cooling and heating. Gas is self-shielded, advects metals, is heated by a homogeneous ultraviolet background, and can cool to 300 K due to metal and molecular line cooling.  Gas flows, shock fronts, and metal disbursement follow self-consistently from this physics.  Observations of molecular clouds \citep[e.g.,][]{KrumholzTan07} are  used to guide the model of star formation.  Star formation occurs in the dense, cold molecular phase ($n_{\hbox{\tiny H}} \sim 100$ {\cc}, $T \simeq 100$ K), which is disrupted by the combination of radiation pressure and photoionisation by massive stars.  The star formation rate is proportional to the gas density divided by the free fall time of the molecular cloud.  These simulations adopted an observationally motivated low (few percent) efficiency per free fall time for converting gas into stars. Runaway young, hot stars are included according to \citet{ceverino09} by providing one third of the newly-formed star particles with a random velocity kick.

In addition to SNe Type Ia and II explosions, photoionisation heating, radiation pressure, and shocked stellar winds from massive stars are incorporated \citep[see][for details]{ceverino09, Ceverino10, Trujillo15, ceverino14}.  The dynamical effect of photoionisation heating is also included by adding a non-thermal pressure $P/k$ (in K cm$^{-3}$), where $10^6 \leq P/k \leq 5\!\times\!10^7$, to the gas surrounding young stars, based on observations of {\HII} regions \citep[e.g.,][]{Lopez14}. This pressure decreases rapidly in order to reproduce the declining density within a growing {\HII} region.  Concurrently, mechanical energy from stellar winds and SN type II thermalises and is injected into the gas as thermal energy around young stars following the rates predicted by Starburst99 \citep{Leitherer99} for the Chabrier IMF. Radiation pressure from young massive stars due to momentum from the radiation field is also included (this momentum couples to the gas and dust through scattering and absorption). UV absorption photons scales as $1 - e^{-\tau_{\hbox{\tiny UV}}}$ and IR scattering scales as $\tau_{\hbox{\tiny IR}} \simeq 1$--10 \citep[e.g.,][]{KrumholzThompson12, Davis14}. 

\begin{table}
\centering
\caption{Simulated Galaxy Properties \label{tab:galaxy}}
\begin{tabular}{r c c c c c c} 
\hline\hline 
Gal ID & $\log(M_{vir})$ & $\log(M_{*})$ & $R_{vir}$ & SFR & log(sSFR) \\[-0.1ex] 
($z=1$) & [M$_{\odot}$] & [M$_{\odot}$] & [kpc] & [M$_{\odot}$~yr$^{-1}$] & [yr$^{-1}$]\\ [0.1ex]
 \hline 
dwALL\_8 &  10.35 & 6.9 & 43.4 & $3\times 10^{-3}$ & $-9.42$\\
dwALL\_1 &  10.36 & 7.3 & 43.7 & $2\times 10^{-2}$ & $-8.99$\\
\hline 
\label{table1}
\end{tabular}
\end{table}

\begin{figure*}
\centering
\includegraphics[width=0.95\hsize]{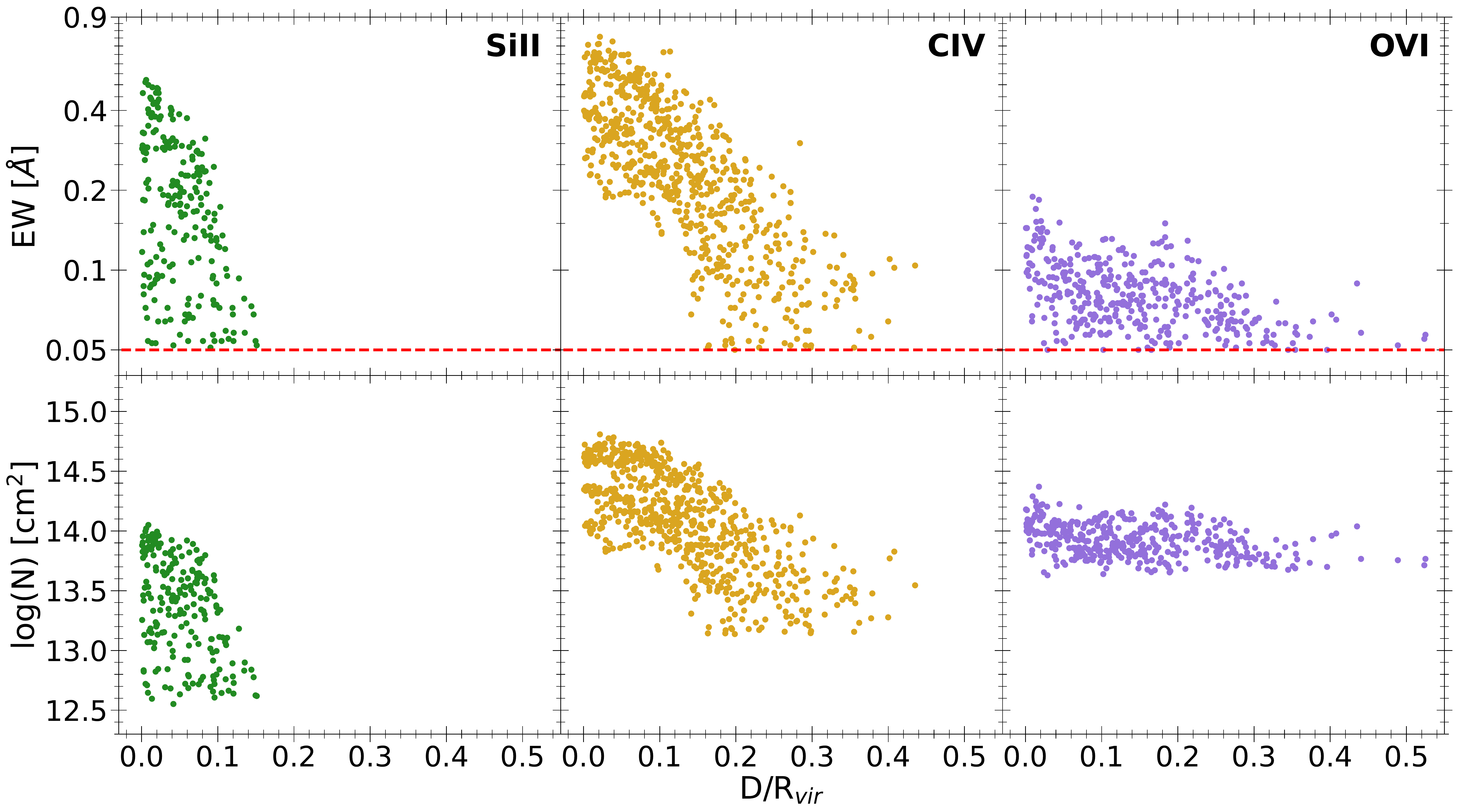}
\caption{(upper) Rest-frame equivalent width (EW) versus $D/R_{vir}$ for the  {\SiII}~$\lambda 1260$, {\CIV}~$\lambda 1548$,  and {\OVI}~$\lambda 1031$ absorption profiles (note the logarithmic scale).  A minimum ``detection threshold'' of ${\rm EW} = 0.05$~{\AA} corresponding to the thresholds of high signal-to-noise high resolution spectra has been applied, as shown by the red dashed line. (lower) The apparent optical depth column density ($\log N/{\rm cm}^{-2}$) for the {\SiII}, {\CIV}, and {\OVI} ions versus $D/R_{vir}$. The synthetic spectra are based on the {\it HST}/COS instrument and have a signal-to-noise ratio of ${\it SNR} = 30$. Lines of sight with no detected absorption or absorption having ${\rm EW} < 0.05$~{\AA} are omitted from this diagram.
}
\label{LOS_props}
\end{figure*}

The dwarf galaxy simulations have high-resolution regions surrounding the galaxies that extend $\sim\! 1$--2 Mpc in diameter.  Each gas cell stores the hydrogen density, $n_{\hbox{\tiny H}}$, temperature, $T$, and metal mass fraction, $x_{\hbox{\tiny M}}$. We extracted smaller post-production boxes centred on the target galaxies that are roughly four virial radii ($4 R_{vir}$) in diameter. The highest gas cell resolution ranges from 20--200~pc, depending on redshift. Within the virial radius of these two $z=1$ galaxies, the median cell resolution is $440$~pc with a median absolute deviation of $320$~pc. The  minimum stellar particle mass of $100$~M$_{\odot}$ and the minimum dark matter particle mass is $9.4 \!\times\! 10^{4}$~M$_{\odot}$. The initial conditions and feedback recipes of the two galaxies are identical, except for the photoionisation heating, where $P/k$ is a factor of eight higher for dwALL\_8 relative to dwALL\_1, which has $P/k = 1 \times 10^6$~K~cm$^{-3}$. These two galaxies were selected with no {\it a priori\/} knowledge of their properties, such as their star formation rates, and stellar or halo masses. Their properties are listed in Table~\ref{tab:galaxy}.






\section{Methodology}
\label{Methodology}

\begin{figure}
\centering
\includegraphics[width=0.95\hsize]{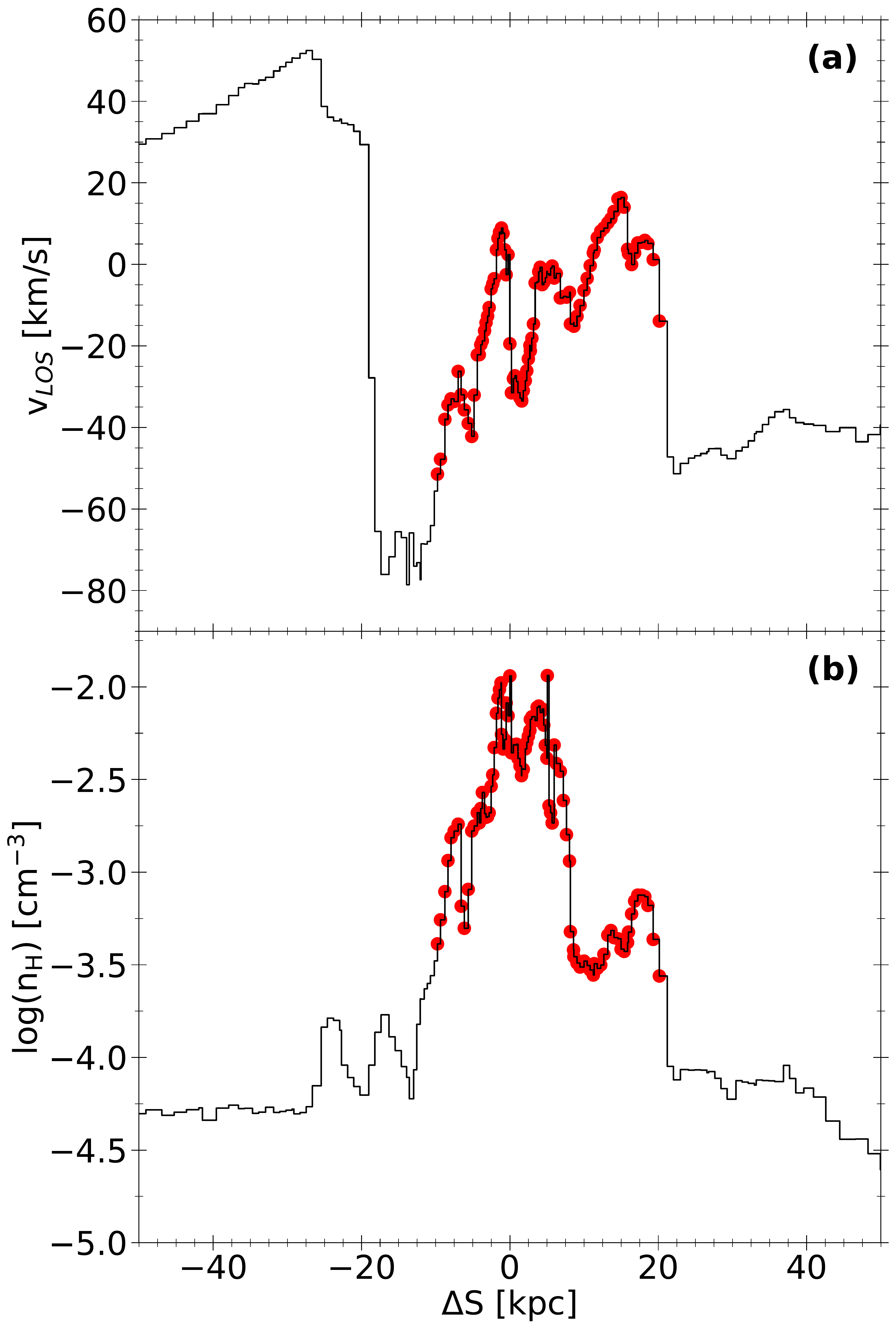}
\caption{Example LOS data. (a) The LOS velocity ($v_{\hbox{\tiny LOS}}$, {\kms}) of the pierced cells as a function of LOS position ($\Delta S$, kpc). The plane of the sky, which slices through the central galaxy, is at $\Delta S=0$~kpc. The observer is toward the left (negative $\Delta S$) and the background quasar is to the right (positive $\Delta S$). (b) The cell hydrogen number density ($n_{\hbox{\tiny H}}$, cm$^{-3}$) as a function of $\Delta S$. The absorbing cells (those which account for 95\% of the rest-frame equivalent width)) for the {\CIV}~$\lambda 1548$ absorption profile for this LOS are marked using red points.}
\label{AbsCellEx}
\end{figure}

For this work we aim to characterise the LOS spatial distribution of absorbing gas ``clouds'' giving rise to each individual {\it absorption component\/} in the observed absorption profiles.  We emphasise that we are not investigating the CGM properties as a function of the sub-grid physics employed in the simulations. We quantitatively assess the correspondence between absorption line components arising from CGM gas structures and the gas structures themselves, i.e., does each absorption component correspond to a unique gas ``cloud?'' If not, how are the gas structures giving rise to a single absorption component spatially distributed along the line of sight?  In this section, we describe how we locate the LOS through the simulated galaxies, generate the mock absorption line spectra, and define ``absorption components'' and ``clouds.'' 

Following the methods described in \citet{marra21a, marra21b}, we generate synthetic (or ``mock'') quasar spectra of the CGM of the simulated galaxies and analyse the resulting absorption profiles. In order to study the low-, intermediate-, and high-ionisation phases that arise in a single absorption line ``system'', we study the absorption arising from the {\SiII}, {\CIV}, and {\OVI} ions.  The Si$^{+}$ has a ground-state ionisation potential of 16.3~eV, which is just higher than that of neutral hydrogen at 13.6~eV. As such, {\SiII} absorption traces low-ionisation conditions found in gas that is dominated by neutral {\HI} absorption. The C$^{+3}$ ion has an ionisation potential of 64.5~eV, which is slightly above that of ionised helium at 54.4~eV, corresponding to an intermediate ionisation condition.  The O$^{+5}$ ion ionises at 138.1~eV. For collisional ionisation this ion peaks at $T\simeq 300,000$~K, and for photoionisation arises when the ratio of the number density of ionising photons to the number density neutral hydrogen is of the order 1 to 10, corresponding to a high-ionisation condition.

Our experimental design and methods for generating the absorption spectra are described in Section~\ref{Mockspec}. As our experiment is focused on the CGM gas that participates in creating (or gives rise to) the objectively detected absorption in the spectra, we identify the individual gas cells probed by a line of sight that collectively are responsible for more than 95\% of the observed equivalent width.  Details are described in Section~\ref{DefAbsCells}. 

\begin{figure*}
\centering
\includegraphics[width=0.95\hsize]{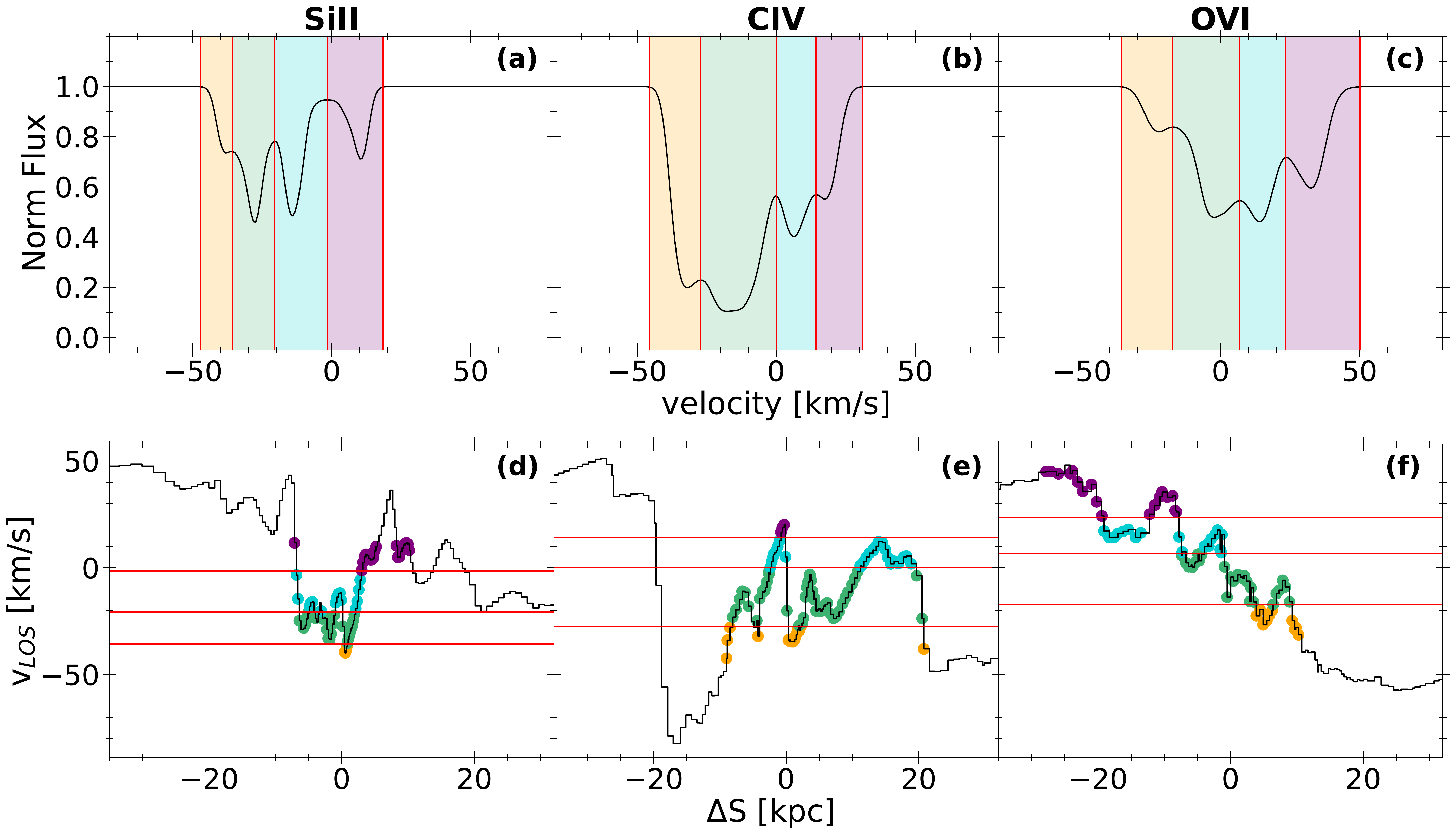}
\caption{(upper) The simulated noiseless spectra of three sample absorption profiles (a) {\SiII}~$\lambda 1260$, (b) {\CIV}~$\lambda 1548$, and (c) {\OVI}~$\lambda 1031$. Each profile is divided into distinct absorption components using the critical and inflection points as described in Section~\ref{AbsCompAnal}. Each absorption component is indicated by the red vertical lines and different colour shading. Each absorption profile is treated in isolation; there is no cross correlation or profile information exchanged between ions. (lower) The gas cell LOS velocity ($v_{\hbox{\tiny LOS}}$, {\kms}) as a function of LOS position ($\Delta S$, kpc) for (d) {\SiII}, (e) {\CIV}, and (f) {\OVI}. The red horizontal lines and the coloured cells correspond to the absorption components shown in (a), (b), and (c), respectively. These examples serve to show that a single absorption component can arise from gas concentrations in multiple non-contiguous LOS positions.}
\label{spec_abs}
\end{figure*}

Metal-line absorption profiles often comprise multiple ``components'' (for example, see Figure~\ref{spec_abs}(a,b,c)). It is common practice that kinematically complex absorption profiles are decomposed into individual Voigt profile (VP) components \citep[e.g.,][]{pb90, pb94, carswell91, cvc03, churchill20, tripp08, werk13, muzahid15, Pointon_2019, haislmaier21}.  As our aim is to examine the physical line-of-sight structure of the absorbing gas associated with each absorption component, we require an objective method for defining each component.  We opt to not employ intensive VP fitting methods, which involve subjective fitting philosophies and are complicated by multi-phase ionisation structure. Instead, we adopt a fast, efficient, and objective method that employs the derivatives and inflection points across noiseless versions of the absorption profiles.  This approach is somewhat similar in principle to those in which the minimum number of VP components are fitted to absorption profiles \citep[e.g.,][]{churchill20}. We describe our method in Section~\ref{AbsCompAnal}.

Finally, in Section~\ref{DefAbsClouds}, we describe our methods for  defining the spatially distinct absorbing gas ``clouds'' that give rise to a given absorption component in the absorption profile of a given ion.

\subsection{Synthetic Spectra and Absorption Line Detection}
\label{Mockspec}

Every gas cell in a hydro{\sc ART} simulation has a unique 3D spatial coordinate defined by its centre position, physical size ($L_{\rm cell}$), 3D velocity components, temperature ($T$), hydrogen number density ($n_{\tH}$), and metal mass fraction ($x_{\tM}$). We perform a post-processing equilibrium ionisation modelling to determine the ionisation fractions in order to calculate the number densities of all ion stages. We employed the photo+collisional ionisation code {\sc Hartrate} \citep[detailed in][]{cwc14}, which defaults to solar abundance mass fractions for each individual metal up to zinc \citep{Draine, Asplund09}.  The {\sc Hartrate} code is well suited for studying the low-density CGM ($\log (n_{\tH}/{\rm cm}^{-3})  <-1$) as it gives the best results for optically thin, low density gas. We note that while the total {\HI} column density for some LOS are optically thick, $\log (N_{\tHI}/{\rm cm}^{-2}) > 17.2$, the post-processing ionisation corrections are conducted on a cell-by-cell basis, and most cells are optically thin\footnote{An examination of the distribution of $\log N_{\tHI}$ of absorbing cells for the LOS incorporated into this study shows all cells have  $\log (N_{\tHI}/{\rm cm}^{-2}) < 16.4$ for {\CIV} and {\OVI} absorbers.  For {\SiII} absorbers, less than 10\% of absorbing cells have  $17.0 \leq \log (N_{\tHI}/{\rm cm}^{-2}) \leq 17.8$.}.  The code does not account for the cumulative column densities of cells locally embedded in higher density neutral regions.

As done in multiple previous studies, \citep{cwc1317b, kcn12, churchill15, Kacprzak_2019, marra21a, marra21b} we have used {\sc Hartrate} for our analysis. A quantitative comparison with the industry-standard ionisation code {\sc Cloudy} \citep{Ferland98,Ferland13} shows that across astrophysically applicable densities and temperatures for optically thin gas, the ionisation fractions are in agreement within $\pm 0.05$ dex. In the post-production boxes, all the gas cells are illuminated with the ultraviolet background (UVB) spectrum of \cite{HaardtMadau2005} to obtain the equilibrium solution. {\sc Hartrate} records the electron density, ionisation and recombination rate coefficients, ionisation fractions, and number densities for each gas cell and for all ions from hydrogen through zinc.

We use the {\sc Mockspec}\footnote{https://github.com/jrvliet/{\sc Mockspec}} pipeline \citep[see][]{churchill15, rachel_thesis} to generate a user-specified number of "quasar" lines of sight (LOS) that are distributed across the ``sky projection'' of the simulated galaxies. Relative to the simulated galaxy, each LOS is defined by the position angle on the plane of the sky (in the range of $0^{\circ} \leq \phi \leq 360^{\circ}$), the impact parameter, and the sky-projected inclination of the galaxy. The plane of the sky is defined as the plane through the centre of mass of the galaxy that is perpendicular to the LOS.  

The details of the methodology to produce synthetic spectra from quasar sightlines and details on how the profiles are synthesised are outlined in \citet{churchill15}.  For this study, we generate spectra having the pixelisation and resolution of COS spectra \citep{green-cos} with a signal-to-noise ratio of ${\it SNR}=30$. We selected this instrumental configuration in order to provide some degree of direct comparison between our work and published observational studies of CGM absorption. As part of the {\sc Mockspec} pipeline, the absorption features are automatically and objectively detected using methods detailed in \cite{churchill00} as originally developed by \cite{schneider93}. The equivalent widths, apparent optical depth column densities \citep[e.g.,][]{savage91}, and a host of additional absorption quantities are also measured automatically using the methods described in Appendix A of \citet{cv01}.  

As we are examining the relationship between absorption profiles and the LOS distribution of the absorbing gas, the 3D orientations of the LOS relative to the sky-projected orientation of the galaxies is not relevant.  We ran {\sc Mockspec} on both galaxies at two orientations for each: face-on and edge-on. For each galaxy at each orientation, we generated 1000 LOS at randomly determined impact parameters (ranging from $\simeq$ 0.5--65~kpc) and position angles, resulting in a total sample of 4000 sightlines. As discussed above, we focused our analysis on absorption from the {\SiII}, {\CIV}, and {\OVI} ions. All detections were at the $5\sigma$ significance level, i.e., ${\rm EW}/\sigma_{\hbox{\tiny EW}} \geq 5$.  For fine-structure doublets, both transitions had to be detected. We include all formally detected absorption profiles for which the {\SiII}~$\lambda 1260$, {\CIV}~$\lambda 1548$,  and {\OVI}~$\lambda 1031$ rest-frame equivalent widths satisfy ${\rm EW} \geq 0.05$~{\AA}. This resulted in a sample of 217 LOS with {\SiII} absorption, 690 LOS with {\CIV} absorption, and 395 LOS with {\OVI} absorption. 

The distributions of equivalent width and column density as a function of impact parameter normalised by the virial radius ($D/R_{vir}$) are shown in Figure~\ref{LOS_props} for each of the ions. As expected, both the equivalent width and column density fall off with increasing $D/R_{vir}$. It is difficult to make direct comparisons with observed distributions as CGM absorption properties of similar mass dwarf galaxies have not been explored at $z=1$.  However, for the {\CIV} absorption, a comparison with COS-Dwarfs galaxies observed at $z\simeq 0$ shows a qualitatively excellent match \citep[see Figure~2 of][]{bordoloi-cosdwarfs}. Our simulated galaxies are reasonable analogues to the COS-Dwarfs galaxies, with specific star formation rates of $\log ({\rm  sSFR/yr}^{-1}) =-9.4$ and $-8.9$ falling within the range $-12.1 \leq \log ({\rm  sSFR/yr}^{-1}) \leq -8.8$ of the COS-Dwarfs sample. 

\subsection{Defining Absorbing Gas Cells}
\label{DefAbsCells}

The {\sc Mockspec} pipeline also automatically determines which gas cells along a given sightline contribute to absorption features.  For gas cell $i$ that is pierced by the LOS, the ionic column density of the target ion $x$ is calculated using the product of the ion's number density ($n_{i,x}$) and the actual path length of the LOS through the cell, i.e.,  not the simple product $n_{i,x}L_{\rm cell}$. If a gas cell has an ion column density of $N_{i,x} < 10^9$~cm$^{-2}$, it is assumed to not contribute to detectable absorption. {\sc Mockspec} determines which of the remaining gas cells contribute to the observed absorption by identifying which gas cells account for 95\% of the equivalent width of the absorption features. That is, individual cells that contribute less than 5\% to the equivalent width are considered to not contribute to the detected absorption. The remaining gas cells are considered to be ``absorbing cells.'' For each LOS, this process is repeated for each ion resulting in a list of absorbing cells for each ion for each LOS.  The absorbing gas cell properties, including LOS position, LOS velocity, gas physical conditions, and cell ID number are stored for later analysis.

In Figure~\ref{AbsCellEx}, we show the velocity of all the {\CIV} gas cells intercepted by an example sightline as a function of line of sight position $\Delta S$ (in kpc), where $\Delta S=0$~kpc is the plane of the sky intersecting the centre of mass of the simulated galaxy. The absorbing cells for {\CIV} are plotted as the red points.

\subsection{Defining Absorption Components and their Gas Cells}
\label{AbsCompAnal}

In order to characterise the LOS spatial distribution of absorbing gas cells for each individual {\it absorption component\/} in detected absorption profiles we developed a method to objectively determine the number and LOS velocity locations of the absorption components comprising the profiles. Using noiseless versions of the profiles, we calculated the discrete first and second derivatives as a function of LOS velocity position across the profiles.  This yielded both the LOS velocities corresponding to the critical points and inflection points along the profiles. The critical points provide the profile extrema and the inflection points determine where the profile concavity changes.  We used the inflection points to bracket the profile minima, thereby identifying the individual absorption components. We used the local maxima between inflection points to define the velocity extent of each absorption component. 

\begin{figure}
\centering
\includegraphics[width=0.95\hsize]{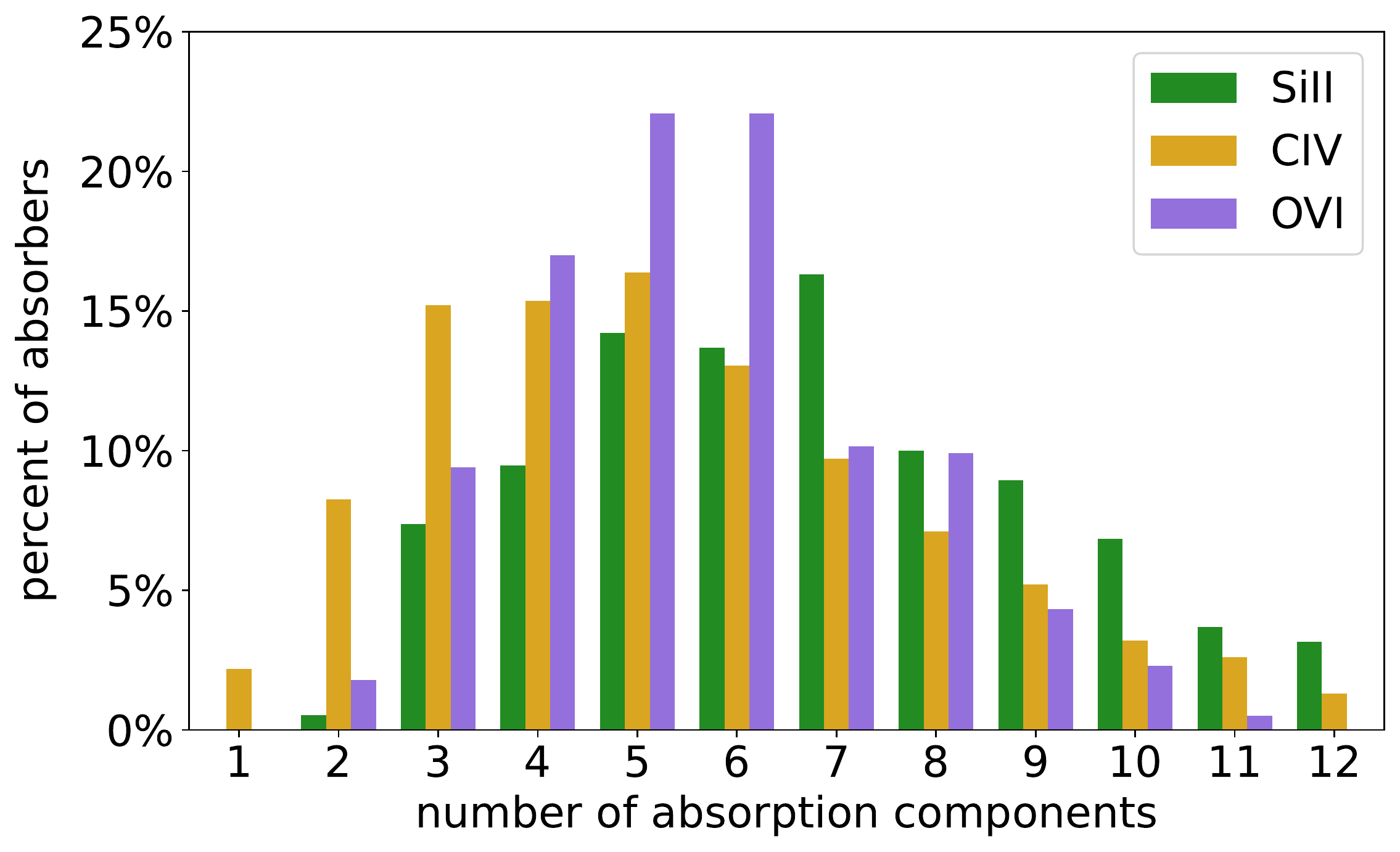}
\caption{The percentage of absorbers having a given number of absorption components for {\SiII} profiles (green), {\CIV} profiles (gold), and {\OVI} profiles (purple).} 
\label{numlos}
\end{figure}

In Figure~\ref{spec_abs}(a,b,c) we provide illustrative examples of absorption components for three selected {\SiII}~$\lambda 1260$, {\CIV}~$\lambda 1548$, and {\OVI}$~\lambda 1031$ absorption profiles, respectively. Red vertical lines mark the velocity limits of each absorption component, which are coloured differently for clarity. Using this technique, we obtained a sample size of 1,787 {\SiII}, 3,723 {\CIV}, and 2,245 {\OVI} unique absorption components. The probability of the number of absorption components for {\SiII}, {\CIV}, and {\OVI} is illustrated in Figure~\ref{numlos}.  Consistent with real-world observations, absorption profiles obtained from simulations are kinematically complex. We find that the ionic transitions we study here have somewhere between four to seven components on average and that the distribution of kinematic complexity is different for each. Our selection criteria for absorption systems has deselected the so-called classical single-cloud weak system \citep[e.g.,][]{weakII}, especially in the low-ionisation phase. Recall that, whereas the absorption components are determined using noiseless spectra (as presented in Figure~\ref{spec_abs}(a,b,c)), the absorption profile detections and properties (equivalent widths, column densities, etc.) are measured on spectra with added Poisson noise with a signal-to-noise ratio of ${\it SNR}=30$.

We next identify the absorbing cells that give rise to each absorption component. Our criterion is that cells with a LOS velocity residing within the LOS velocity range of the absorption component are identified with the absorption component. Shown in Figure~\ref{spec_abs}(d,e,f) are the LOS velocities versus LOS position for the cells along the lines of sight. The red horizontal lines provide the LOS velocity ranges of the absorption components; these lines correspond to the vertical red lines in the corresponding absorption profiles shown in Figure~\ref{spec_abs}(a,b,c) based on the critical points in the absorption profiles. For illustration purposes,  we coloured the absorbing cells to correspond to the colour shading of their corresponding absorption components.

Due to the complexity of LOS velocity flows along LOS position, an absorption component can map to multiple spatially separated groupings of absorbing cells. For example, for the {\CIV}~$\lambda 1548$ profile illustrated in Figure~\ref{spec_abs}(b,e), the absorption component centred at $v_{\hbox{\tiny LOS}} \simeq 8$~{\kms}  (coloured turquoise) arises from two groupings, one spanning the LOS range $-3 \leq \Delta S \leq 0$~kpc and a second with $10 \leq \Delta S \leq 20$~kpc. Similarly, for the {\OVI}~$\lambda 1031$ profile illustrated in Figure~\ref{spec_abs}(c,f), the absorption component with $v_{\hbox{\tiny LOS}} \simeq 32$~{\kms} (coloured purple) arises from two groupings with $-28 \leq \Delta S \leq -18$~kpc and a second with $-12 \leq \Delta S \leq -7$~kpc. In the following section, we quantify these groupings by defining ``absorption component clouds.''

\begin{figure*}
\centering
\includegraphics[width=0.95\hsize]{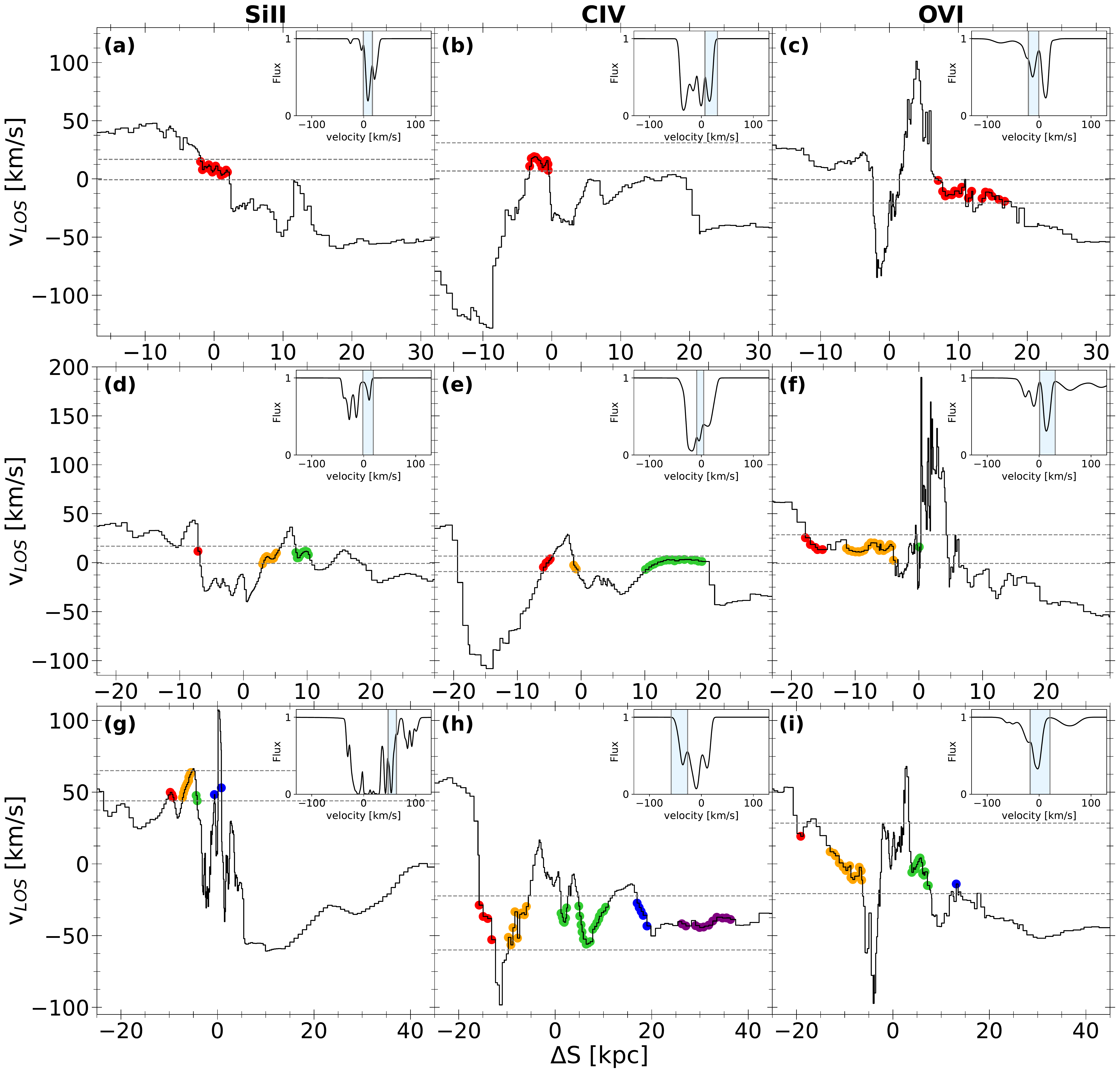}
\caption{Velocity as a function of position along the LOS ($\Delta S$) for an absorption component with a single, few, or multiple spatially distinct gas clouds for (a,d,g) {\SiII}, (b,e,h) {\CIV}, and (c,f,i) {\OVI}. (a,b,c) All of the gas for that absorption component is in a single gas cloud. (d,e,f) The gas for the absorption component is separated into 3 spatially distinct clouds. (g,h,i) The gas for the absorption component is separated into 4 or 5 spatially distinct clouds. The black line shows the velocity as a function of $\Delta S$ for all gas cells the LOS intersects. The coloured points represent cells that give rise to absorption, and the different colours corresponds to spatially distinct clouds. The inset plots show the spectrum for that ion, where the shaded grey region highlights the associated absorption component. }
\label{spec_diff_comps}
\end{figure*}

\subsection{Defining Absorption Component ``Clouds''}
\label{DefAbsClouds}

When analysing absorption profiles, astronomers typically engage in modelling techniques founded on the assumption that each absorption component corresponds to a single cloud-like gas structure. We aim to investigate if this assumption is supported by studies of hydrodynamics simulations, where the absorbing gas structures can be directly examined. Here, we describe how we define an absorbing ``cloud'' directly associated with the absorption component to which it gives rise in a mock spectrum.

Ideally, an absorbing cloud is defined as a series of spatially contiguous absorbing gas cells clustered along the line of sight in a range of $\Delta S$ LOS positions. Furthermore the Eulerian adaptive mesh implemented for the hydrodynamics, the gas cells along the LOS vary in size, $L_{\rm cell}$.  This results in a staggered and variable alignment of cells in three dimensions. Further, the LOS orientation through the simulation cube is not aligned with the cube principle axes, as the LOS  are defined relative to the galaxy orientation in the cube.  For these reasons, there is no fixed cell-to-cell LOS separation ($\Delta S_{n} - \Delta S_{n+1}|$) between cell $n$ and the sequential cell $n+1$ along the LOS. Therefore, we adopt a method in which we account for both the average cell size $\langle L_{\rm cell}\rangle$ of the absorbing cells and the staggered alignment of their positions. 

We scan the absorbing cells in the velocity range of a given absorption component in the direction of increasing LOS position $\Delta S$ and compute  
\begin{equation}
R_n = \frac{|\Delta S_{n} - \Delta S_{n+1}|}{\langle L_{\rm cell}\rangle} \, , 
\label{distance}
\end{equation}
where $\Delta S_{n}$ corresponds to the $n$th absorbing cell along the LOS, $\Delta S_{n+1}$ is the absorbing cell with the next highest $\Delta S$ position, and $\langle L_{\rm cell}\rangle$ is the average cell size of the absorbing cells {\it in the velocity range of the absorption component}. We then adopt the criterion that if $R_n\leq R_{\rm crit}$, the two cells, $n$ and $n+1$, are grouped into an absorbing cloud. After exploration of $R_{\rm crit}$ in the range 1 to 10, we adopted $R_{\rm crit} = 5$.  When $R_n >  R_{\rm crit}$ then cells $n$ and $n+1$ are segregated into separate clouds. As we discuss below, this method exhibits flexibility in challenging cases and robustly identifies ``clouds'' even if they have large velocity shears across the LOS. As defined in Section~\ref{DefAbsCells}, an absorbing cell is not simply defined as one that resides in the velocity range of an absorption profile, but must also contribute at least 5\% to the measured rest-frame equivalent width of the profile. Further, only absorbing cells are included in the implementation of Eq.~\ref{distance}.

In Figure~\ref{spec_diff_comps}, we illustrate several examples for which implementation of Eq.~\ref{distance} yields different numbers of spatially distinct absorbing gas clouds for selected absorption components. Each panel shows the LOS velocity $v_{\hbox{\tiny LOS}}$ versus LOS position $\Delta S$ for all gas cells intersecting the LOS. The inset plots in the upper right corner of each panel show the absorption profile for the given ion with the selected absorption component shaded. In the main panels, the velocity range of the selected absorption component is shown as horizontal lines and the absorbing gas cells giving rise to the selected absorption component are colour coded.

In Figure~\ref{spec_diff_comps}(a,b,c), we show example LOS for {\SiII}, {\CIV}, and {\OVI} absorption for which the selected absorption component has a single spatially isolated cloud (the red coloured cells) giving rise to the absorption. In Figure~\ref{spec_diff_comps}(d,e,f), we show example LOS for which three spatially distinct clouds (coloured red, orange, and green) contribute to the selected absorption component. In Figure~\ref{spec_diff_comps}(g,h,i), we show example LOS for which four (panels g,i) and five (panel h) spatially distinct clouds (coloured red, orange, green, blue, and purple) contribute to the selected absorption component.

We make three observations regarding the definition of absorbing clouds.  First, as seen in Figure~\ref{spec_diff_comps}(c), non-contiguous cells can be assigned to a single cloud. In this case, the three cells excluded from the cloud (the cells at $\Delta S \simeq +12$~kpc) do not reside in the velocity range of the absorption component. As the velocity range of an absorption component is not a perfect definition devoid of some unquantifiable degree of ambiguity, there is some ``fuzziness'' that might be termed as edge effects. In this example, we see that our method accounts for this fuzziness by allowing the three cell discontinuity in the assigned cloud. Second, as seen in Figure~\ref{spec_diff_comps}(g) at $\Delta S \simeq 0$~kpc, there is a rapid LOS velocity inversion in which a single cell on each side of the inversion resides in the velocity range of the absorption component. Intuitively, these two cells are part of a dynamically coherent structure spanning a LOS velocity range of $\sim\! 100$~{\kms} even though they are not in contiguous cells along the LOS path. Our method has captured these coherent and dynamic structures such that the cells embedded in such structures are grouped into a single cloud.  This brings us to our third point,  which is that some clouds can be single cells, as seen in Figure~\ref{spec_diff_comps}(d) at $\Delta S \simeq -7$~kpc. Though our method is not the only possible approach to defining absorbing clouds, it captures important features and exemplifies flexibility when faced with challenging and ``fuzzy'' scenarios.

\section{Results}
\label{Results}

\begin{figure}
\centering
\includegraphics[width=0.95\hsize]{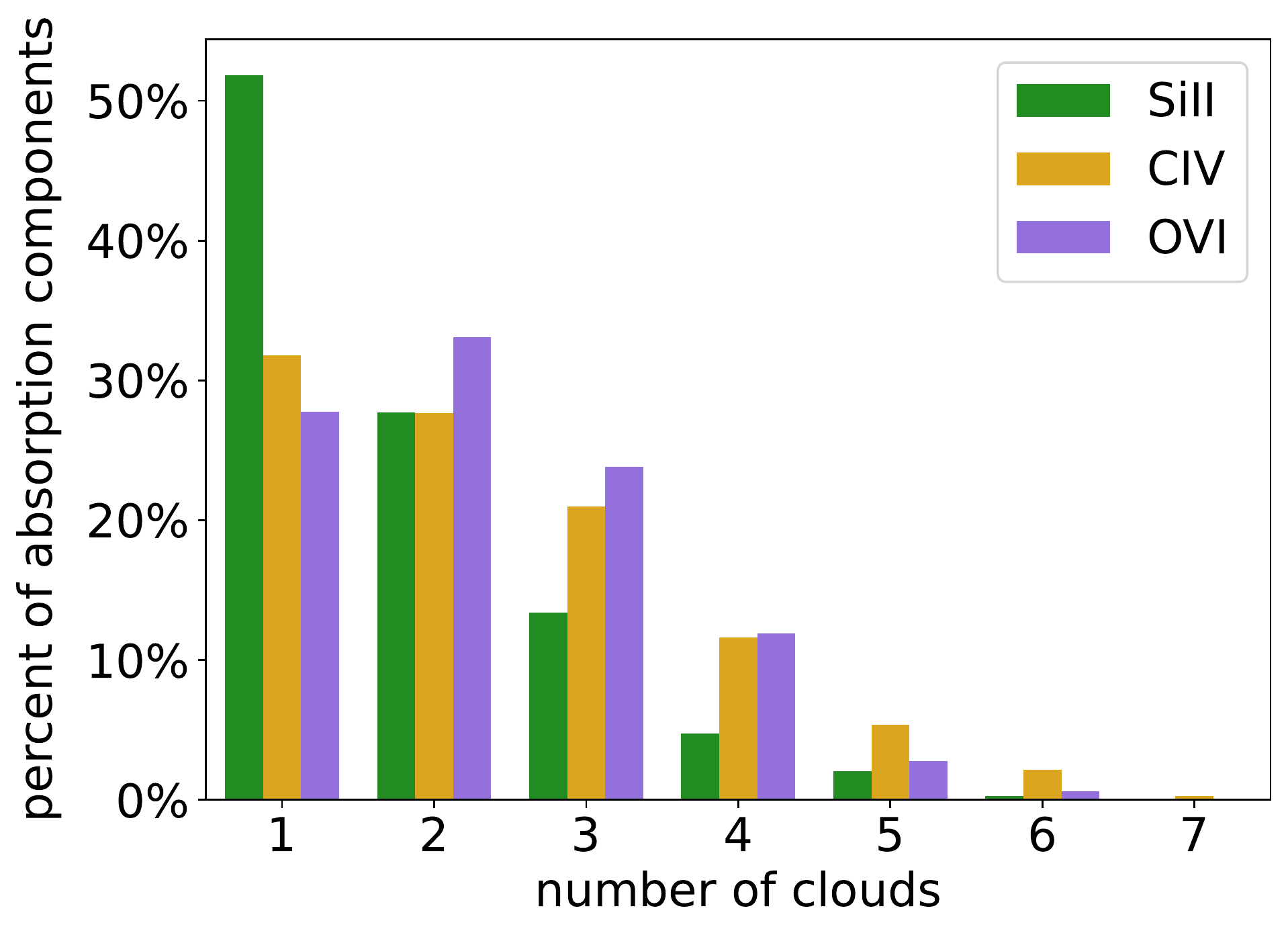}
\caption{The percentage of absorption components having a given number of spatially distinct clouds for {\SiII} absorption components (green), {\CIV} absorption components (gold), and {\OVI} absorption components (purple).} 
\label{numcomps}
\end{figure}

\begin{figure*}
\centering
\includegraphics[width=0.95\hsize]{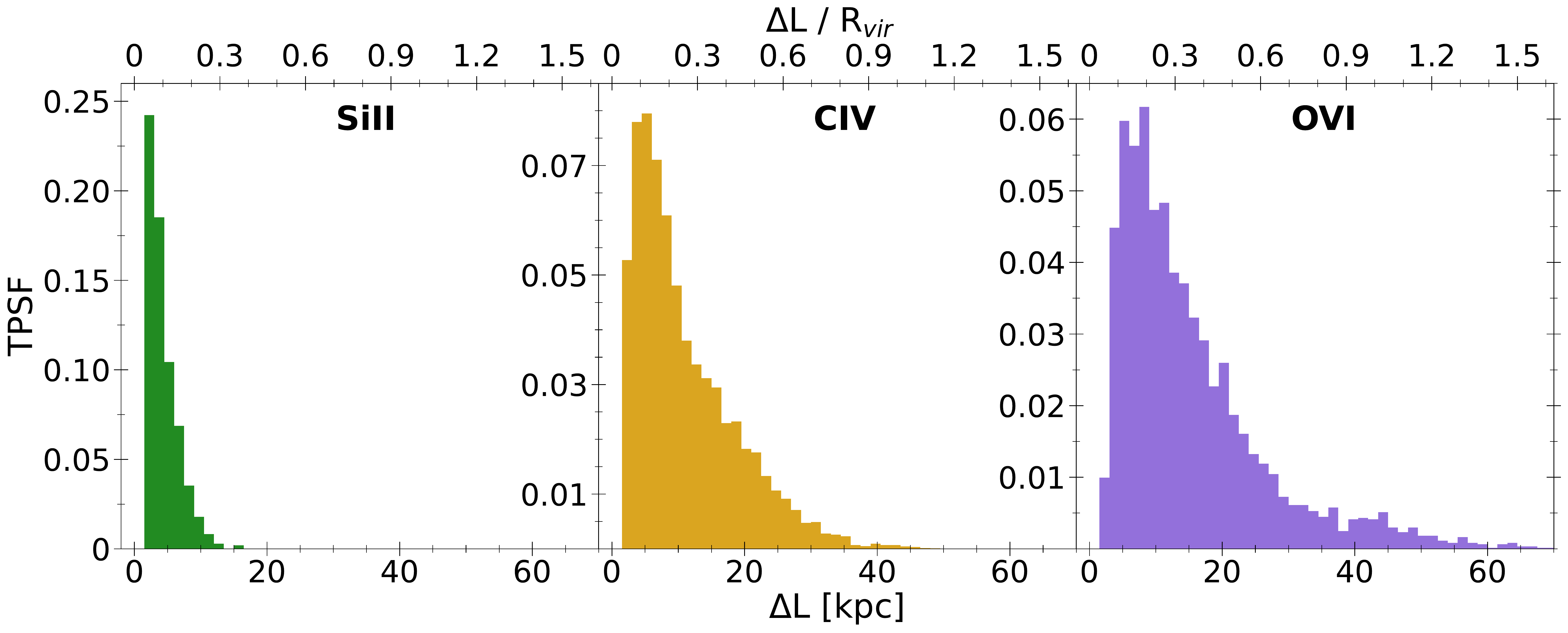}
\caption{The absorbing cloud two-point separation function (TPSF) giving the distribution of pair-wise LOS separations ($\Delta L$) between the spatially distinct gas clouds associated with an absorption component. (left, green) {\SiII} components. (centre, gold) {\CIV} components. (right, purple) {\OVI} components.}  
\label{freq_hist}
\end{figure*}

From our 4000 lines of sight, we located 1,302 that had detectable {\SiII}~$\lambda 1260$, {\CIVdblt}, and/or {\OVIdblt} absorption (217 LOS with {\SiII}, 690 LOS with {\CIV}, and 395 LOS with {\OVI}). Here, we define {\it detectable absorption\/} as absorption objectively identified and determined to have rest-frame ${\rm EW} \geq 0.05$~{\AA} with a significance level of $5\sigma$ or greater, i.e., $\sigma_{\hbox{\tiny EW}}/{\rm EW} \geq 5$. Using an objective methodology based on the overall absorption profile morphology, we split each absorption profile into multiple absorption components.  This yielded a total of 7,755 absorption components (1,787 {\SiII} components, 3,723 {\CIV} components, and 2,245 {\OVI} components). For each absorption component, we located the absorbing gas cells along the line of sight giving rise to that absorption component, thus determining the LOS positions and number of spatially distinct gas clouds that contribute to the absorption component. 

\subsection{The Distribution of Clouds per Absorption Component}

To characterise the distribution in the number of gas clouds giving rise to a single absorption component, we computed the probability mass functions (PMFs), i.e., the discrete distributions of the fraction of absorption components associated with a given number of spatially distinct clouds.   In Figure~\ref{numcomps}, we show the PMFs for {\SiII} (black bars), {\CIV} (green bars), and {\OVI} (blue bars).  Examples of single cloud absorption components are shown in Figure~\ref{spec_diff_comps}(a, b, c) and 3-cloud absorption components in Figure~\ref{spec_diff_comps}(d, e, f). Similarly, an example of a 4-cloud absorption component is shown in Figure~\ref{spec_diff_comps}(h) and of a two 5-cloud absorption component in Figure~\ref{spec_diff_comps}(g,i). No absorption component has more than 7 spatially distinct clouds, and only {\CIV} has more than 6 distinct clouds in an absorption component.  We tabulate the PMFs in Table~\ref{tab:pmf}.

We find that most of the absorption components in {\SiII}~$\lambda 1260$ and {\CIV}~$\lambda 1548$ profiles arise from a single, spatially isolated cloud. {\SiII} absorption components arise from a single cloud about 52\% of the time; there is rapid drop to 28\% for two distinct clouds, and then a drop to 13\% with three distinct clouds. Less than 8\% of all {\SiII} absorption components arise from four or more clouds.  For {\CIV}, somewhat less than a third (32\%) of absorption components arise from a single isolated cloud. Furthermore, the distribution of the number of clouds is flatter than for {\SiII}, with 28\%, 21\%, and 12\% of components arising from two, three, and four distinct clouds, respectively.  On the other hand, a third (33\%) of {\OVI}~$\lambda 1031$ absorption components arise from two distinct clouds, though 28\% arise from a single isolated gas cloud.  There is a 24\% chance that an {\OVI} absorption component arises from four clouds, and a 12\% chance it arises from five clouds.  Our results clearly show that individual absorption components, especially for higher ionisation levels, arise from multiple spatially separated clouds within the simulations.

\begin{table}
\centering
\caption{Distributions of Clouds per Absorption Component \label{tab:pmf}}
\begin{tabular}{c r r r} 
\hline\hline 
 $N_{\rm clouds}$ & {\SiII} & {\CIV} & {\OVI} \\[0.1ex] 
\hline 
1 & 52\%     & 32\%  & 28\%   \\ 
2 & 28\%     & 28\%  & 33\%    \\
3 & 13\%     & 21\%  & 24\%   \\
4 & 5\%      & 12\%  & 12\%    \\
5 & 2\%      & 5\%   & 3\%   \\
6 & 0.3\%    & 2\%   & 0.1\%   \\
7 & 0\%    & 0.3\%   & 0\%   \\
\hline 
\end{tabular}
\end{table}


\subsection{The LOS Cloud-Cloud Separation Clustering Function} 

In order to characterise the distribution of LOS physical separation of the clouds giving rise to a given absorption component, we constructed the absorbing cloud two-point separation function (TPSF). For each component for a given ion, we determined the ion number density weighted mean LOS position of each gas cloud, 
\begin{equation}
  \Delta S_{\hbox{\tiny C}}({\rm X}) = 
  \frac
  {\sum_i n_i ({\rm X}) \Delta S_i}  
  {\sum_i n_i({\rm X})}  \, ,
\label{eq:DScenter}
\end{equation}
where $n_i ({\rm X})$ is the number density of ion X and $\Delta S_i$ is the LOS position of cell $i$, respectively, and where the sum is over the absorbing cells comprising the cloud for ion X. For each component, we then computed the pair-wise differences, $\Delta L$, of the $\Delta S_{\hbox{\tiny C}}({\rm X})$ for that LOS. 

In Figure~\ref{freq_hist}, we show the absorbing cloud TPSF for {\SiII} (black bars), {\CIV} (green bars), and {\OVI} (blue bars). The distribution of two-point cloud separations are shown as a function of both LOS separation $\Delta L$ (in kpc) and the ratio $\Delta L/R_{vir}$. The TPSFs are area normalised and are thus probability distribution functions. In Table~\ref{tab:tpsf}, we list selected statistical descriptors of the TPSFs, including the mode (peak value), and $\Delta L_{50}$ and $\Delta L_{90}$.  The latter two provide the 50\% and 90\% enclosed areas of the PDFs.

When examining absorption components having two or more spatially distinct clouds, we obtain insight into the LOS spatial clustering. For absorption components of {\SiII}, the peak LOS separation of clouds is $\Delta L$ = 1.3~kpc (0.03$R_{vir}$) The frequency of LOS separations tends to drop off such that roughly 50\% of these clouds will have separations less than $\sim\!3$~kpc (0.07$R_{vir}$) and 90\% will have separations less than $\sim\!7$~kpc (0.16$R_{vir}$). For these simulated galaxies, the maximum LOS separation between clouds is approximately 15 kpc, or 0.4 R$_{vir}$. For absorption components of {\CIV}, we see a peak LOS separation at $\Delta L \simeq 5$~kpc (0.1$R_{vir}$) with 50\% clustering within $\sim\!9$~kpc ($0.2R_{vir}$) and 90\% within $\sim\!20$~kpc ($0.5R_{vir}$).  For absorption components of {\OVI}, the most frequent LOS separation is $\Delta L$ = 6~kpc (0.13$R_{vir}$) with 50\% clustering within $\sim\!12$~kpc ($0.3R_{vir}$) and 90\% within $\sim\!32$~kpc ($0.75R_{vir}$).  The clouds of a given {\CIV} absorption component can have separation up to 45 kpc or $1R_{vir}$ and the clouds for a given {\OVI} absorption component can be separated by as much as $\Delta L$ = 100 kpc or about $2.25R_{vir}$.  

The above values may reflect the particular galaxies selected from our simulations, which are low-mass dwarf galaxies with low star formation rates. Though the statistical descriptors and characteristics of the PMFs of the number of clouds per absorption component and the TPSFs of these clouds may have some dependence on galaxy properties such as halo mass and star formation rate, the key result that a single absorption component in a complex absorption profile can arise from multiple physically distinct and highly separated clouds is likely independent of galaxy properties. Exploration of how galaxy properties and stellar feedback recipes affect the PMFs and TPSFs is beyond the scope of the present investigation.

\begin{table}
\centering
\caption{Cloud Two-Point Separation Function Characteristics \label{tab:tpsf}}
\begin{tabular}{c r r r} 
\hline\hline
Descriptor & {\SiII} & {\CIV} & {\OVI} \\[0.1ex] 
\hline 
mode (kpc)             & 1.3     & 5.1   &   6.1 \\ 
$\Delta L_{50}$ (kpc)  & 3.0     & 8.7   & 12.3  \\
$\Delta L_{90}$ (kpc)  & 7.0     & 21.3  & 32.3  \\ 
\hline
mode ($R_{vir}$)        & 0.03    & 0.11  & 0.13 \\ 
$\Delta L_{50}$ ($R_{vir}$) & 0.07    & 0.20  & 0.28 \\
$\Delta L_{90}$ ($R_{vir}$) & 0.16    & 0.49  & 0.74 \\
\hline 
\label{table3}
\end{tabular}
\end{table}

\subsection{The Ion-to-Ion Common-Mass Distribution in Clouds}

\begin{figure}
\centering
\includegraphics[width=0.95\hsize]{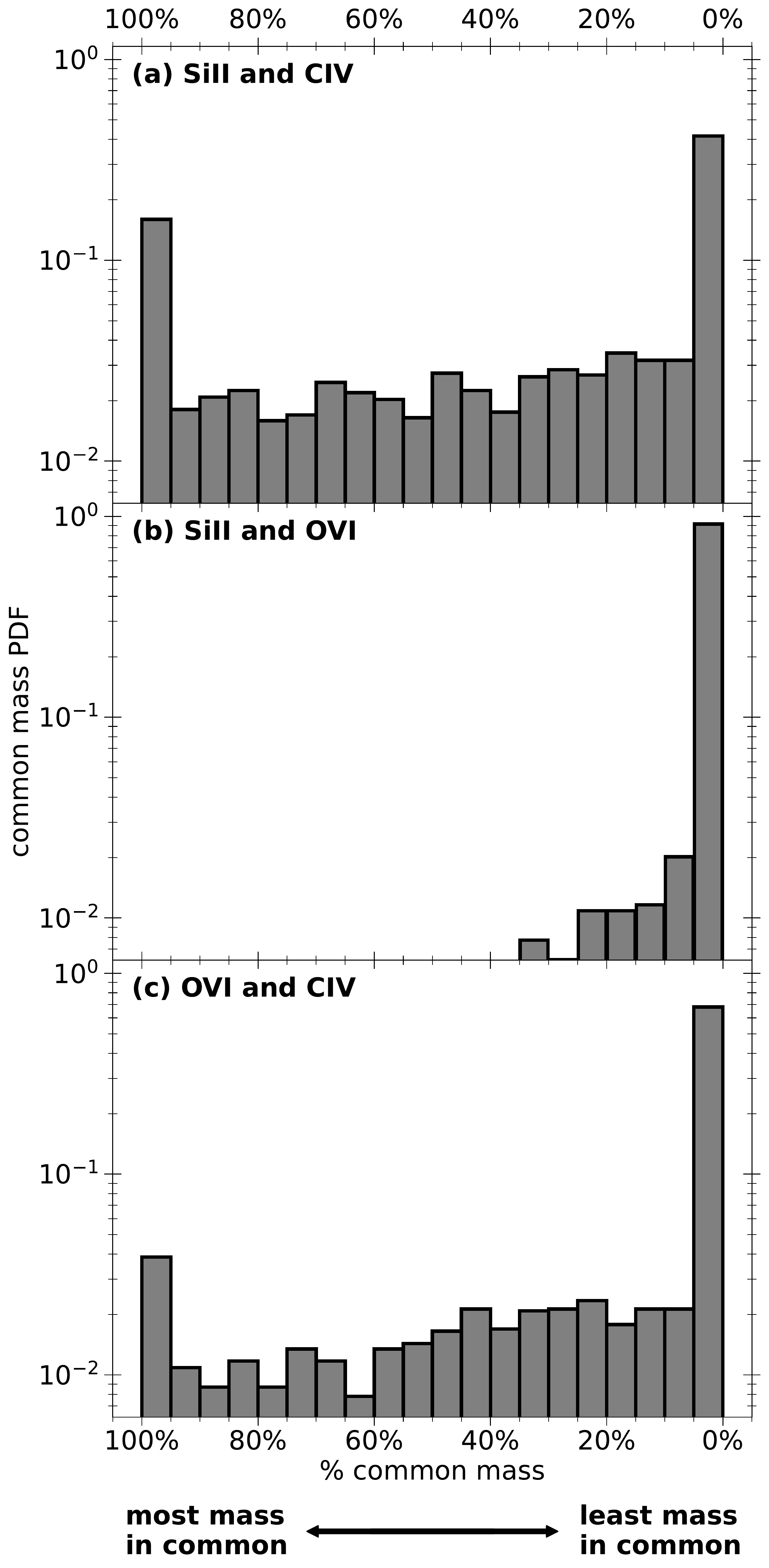}
\caption{The absorption component common mass PDF, which measures the probability that two ions with a velocity-aligned absorption component will have a given percentage of absorbing gas mass in common. (a) {\SiII} and {\CIV}. (b) {\SiII} and {\OVI}. (c) {\CIV} and {\OVI}.  The percentage of common mass is provided in bins of 5\%.  The lower the percent common mass, the more spatially distinct (or physically separated along the line of sight) is the absorbing gas for the two ions.
}
\label{colocated}
\end{figure}

When analysing absorption profiles, it is not uncommon to assume the absorption components from different ions that are aligned in LOS velocity arise from the same gas cloud.  That is, a single phase of gas is assumed to explain the relative strength of the aligned absorption components. We aim to investigate if this assumption is supported by studies of hydrodynamics simulations, where the absorbing gas structures can be directly examined. 

Our methodology for locating absorption components and their associated absorbing clouds has been performed on an ion-by-ion basis.  So far, we have not attempted to couple the clouds associated with the absorption component in one ion with the clouds associated with a LOS velocity aligned absorption component in another ion.  In fact, the manner in which we have independently defined absorption components in a given ion's absorption profile does not naturally yield LOS velocity aligned absorption components in another ion's absorption profile (as would be the case, for example, using VP fitting). Even under these considerations, there can be considerable LOS velocity overlap in the individual components defined for different ions in the same LOS (i.e, ``absorption system'').  We adopt the definition that two individual components for different ions have LOS velocity overlap when 90\% of the union set of their absorbing gas cells reside within the velocity ranges of the two individual components. 

Here, we investigate the question of how much absorbing gas is in common for the clouds associated with a LOS velocity aligned component in two different ions. Consider the clouds associated with the LOS velocity aligned components from ions A and B. Some portions of the clouds associated with A's absorption component may spatially reside in the same LOS location as do some clouds associated with B's absorption component. Conversely, some portions of the clouds associated with A's absorption component may {\it not\/} spatially reside in the same LOS location as some of the clouds associated with B's absorption component. 

We quantify the amount of gas in common to the LOS velocity aligned absorption components of ion A and B as the fraction of gas mass residing in the absorbing cells that are in common to the two ions.  That is, we computed
\begin{equation}
    f_{\hbox{\tiny CM}} = 
    {\displaystyle \sum_{A \cap B}\!  m_i} \, \Bigg/ \,
    \left( {\displaystyle \sum _{A} m_i + \sum_{B} m_i} \right) 
    \, , \quad m_i = m_{\hbox{\tiny H}} \left( \frac{n_{\hbox{\tiny H}}}{x_{\hbox{\tiny H}}} L^3 \right) _i
\, ,
\label{eq:fCM}
\end{equation}
where the cell mass $m_i$ depends on the cell hydrogen number density ($n_{\hbox{\tiny H}}$), hydrogen mass fraction\footnote{Employing $x_{\hbox{\tiny H}} +x_{\hbox{\tiny He}} + x_{\hbox{\tiny M}} = 1$ for each cell, we write $x_{\hbox{\tiny H}} = (1 - x_{\hbox{\tiny M}})/(1 + r)$, where $r = x_{\hbox{\tiny He}}/x_{\hbox{\tiny H}}$. For a primordial gas $r = 0.3334$ and for solar abundances $r=0.3366$ \citep{Lodder19}, we have a range that results in a 0.2\% difference in cell masses. We adopt the average value of $r = 0.335$ for all cells, which is appropriate for metal mass fractions of 0.1--0.01 solar. Our treatment renders $x_{\hbox{\tiny H}}$ to be a function of the mass fraction of metals, $x_{\hbox{\tiny M}}$, in the cell.} ($x_{\hbox{\tiny H}}$), and cell size ($L$). The sum in the numerator is taken over the absorbing gas cells $i$ that are the union of cells for the LOS velocity aligned components of ions A and B. The sums in the denominator are independently taken over the LOS velocity aligned components of ions A and B. As shown in Figure~\ref{spec_diff_comps}, it is possible to have multiple spatially separated absorbing gas clouds contributing to a single absorption component; thus, the $f_{\hbox{\tiny CM}}$ should be interpreted considering this fact.

\begin{table}
\centering
\caption{Percentage of Components with Common Gas Mass$^{\rm a}$ \label{tab:common-massPDF}}
\begin{tabular}{c c c c c} 
\hline\hline 
(1) & (2) & (3) & (4) & (5) \\[0.1ex] 
 ions & $\geq25\%$ & $\geq50\%$ & $\geq75\%$ & $\geq90\%$\\[0.1ex] 
\hline
{\SiII} and {\CIV} & 46\%   & 34\%  & 24\%  & 18\%   \\
{\SiII} and {\OVI} & 3\%  & 0.8\%  & 0.4\%  & 0.2\%    \\ 
{\OVI} and {\CIV}  & 24\%  & 14\%   & 8\%  & 5\%  \\
\hline 
\multicolumn{5}{l}{(a) The entries in this table give the percentage of LOS velocity}\\
\multicolumn{5}{l}{aligned absorption components from two different ions that have}\\
\multicolumn{5}{l}{(col 2) $\geq 25\%$, (col 3) $\geq 50\%$, (col 4) $\geq 75\%$, and (col 5) $\geq 90\%$}\\
\multicolumn{5}{l}{of their gas mass in a common gas structure.}
\end{tabular}
\end{table}

In Figure~\ref{colocated}(a,b,c), we show the PDFs of the absorption component common-mass fraction for {\SiII} and {\CIV}, for {\SiII} and {\OVI}, and for {\OVI} and {\CIV}, respectively.  In Table~\ref{tab:common-massPDF}, we provide the cumulative percentages for which the absorbing clouds associated with the LOS velocity aligned components of two ions have greater than 25\%, 50\%, 75\%, and 90\% of their mass in common. We find that the LOS velocity aligned absorption components of {\SiII} and {\CIV} ions have the highest percentage of mass in common in that 46\% of their absorbing clouds have more than 25\% of their mass in common, 34\% of their absorbing clouds have more than 50\% of their mass in common, and 18\% of their absorbing clouds have 90\% of their mass in common.  {\SiII} and {\OVI} have the lowest percentage of mass in common in that only 3\% of their absorbing clouds have more than 25\% of their mass in common and a scant 0.2\% of their absorbing clouds have 90\% of their mass in common. {\OVI} and {\CIV} also have a lower percentage of mass in common compared to {\SiII} and {\CIV}.  Roughly 24\% of their absorbing clouds have more than 25\% of their mass in common, 14\% of their absorbing clouds have more than 50\% of their mass in common, and 5\% of their absorbing clouds have 90\% of their mass in common.  Overall, our analysis shows that, even though components from different ions align in velocity space within the absorption profiles, there is very little gas mass associated with the individual ions that is in common.
 
\begin{figure*}
\centering
\includegraphics[width=0.8\hsize]{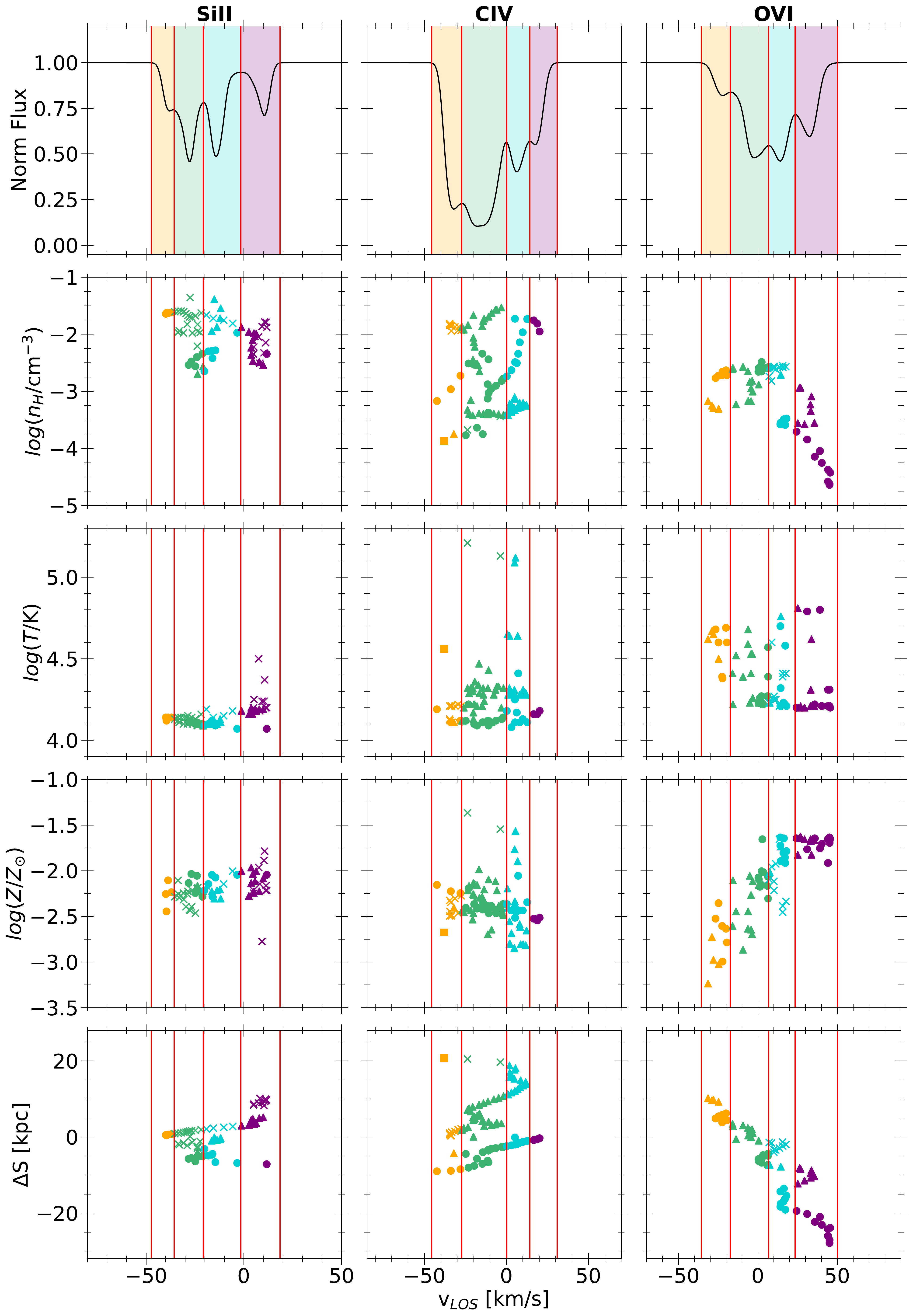}
\caption{The physical conditions of the absorbing gas cells (defined in Section~\ref{DefAbsCells}) as a function of LOS velocity for the absorption profiles also illustrated in Figure~\ref{spec_abs}.  (top row) The simulated noiseless {\SiII}~$\lambda 1260$, {\CIV}~$\lambda 1548$, and {\OVI}~$\lambda 1031$ absorption profiles. Red vertical lines and colour shading indicate distinct absorption components. (second row) The absorbing cell hydrogen number density, $n_{\tH}$, coloured according to its absorption component. (third row) The absorbing cell temperature, $T$. (fourth row) The absorbing cell metallicity, $Z/Z_{\odot}$. (bottom row) The absorbing cell LOS position, $\Delta S$. For a given absorption component, the absorption cells segregate into spatially isolated clouds by their $\Delta S$ separations.  Point types designate each cloud for each component, where circles ({\large $\bullet$}), triangles ($\blacktriangle$), crosses ($\times$), and squares ($\blacksquare$) are plotted in order of increasing $\Delta S$. Each absorption component shown here has one to four clouds.}
\label{AbsCellProps}
\end{figure*}

\section{Discussion}
\label{Discussion}

The most common method for studying the CGM is by analysing CGM absorption lines observed in absorption spectra.  Typically, the first analysis step is to measure ion column densities in individual VP components using Voigt profile fitting software \citep[e.g.,][]{carswell91, welty91, fontana95, mar95, cwc-thesis, vpfit, howarth15, bainbridge17, gaikwad17, liang17, krogager18, churchill20} or by applying the apparent optical depth method \citep[AOD, e.g.,][]{savage91} to obtain the total column density across the full absorption profile.  While VP fitting naturally results in individual kinematically segregated absorption components, it is challenging to separate complex absorption profiles into kinematic components using AOD methods \citep[however, see][]{churchill03, prochaska03, Fox05, Kacprzak12a}. The second analysis step is to apply chemical-ionisation modelling to determine the absorbing gas density, temperature, metallicity, and possibly abundance pattern, using the VP or AOD column densities as the constraints on the best-fit models \citep[e.g.][]{bergeron86, steidel90, cvc03, Fox05, fox15, tripp08, crighton13, werk14, fumagalli16,  glidden16, prochaska17, lehner19, Pointon_2019, Wotta19, haislmaier21}. Notably, while there are a range of ionisation conditions that can be modelled \citep[e.g.,][]{Ferland98, Ferland13, Ferland17, sutherland93, GnatSternberg07, Groves08}, almost all analyses are founded on the assumption that each VP or AOD component maps to a single, spatially isolated cloud.  However, VP component-by-component multi-phase ionisation modelling using the full absorption profile {\it shapes\/} and {\it kinematic structure\/} to constrain the model parameters is continuously being refined \citep[e.g.,][]{charlton00, charlton03, ding03a, ding03b, zonak04, ding05, masiero05, LynchCharlton07, rosenwasser18, Sameer21, Sameer22, nielsen22}. 

Our collective understanding of the metallicities and spatial distribution of the CGM as a function of galaxy property, projected distance, and redshift are based on these analyses. As such, they are a very important pillar on which our astrophysical knowledge is founded. As shown in Figure~\ref{numcomps}, the simulations indicated that more than 50\% of individual absorption components  arise from multiple physically separated gas ``clouds'' having similar LOS velocity (also see Table~\ref{tab:pmf}).  If the results we presented in Section~\ref{Results} reflect the underlying reality of how absorption lines record the CGM gas, then they place tension on current observational analysis methods as they suggest that component-by-component absorption line formation is more complex and convoluted than is assumed and applied for chemical-ionisation modelling.  Further, the common mass PDF shown in Figure~\ref{colocated} indicates that there can be very little overlap in the gas cloud structures giving rise to absorption components from ions having different ionisation thresholds.

In particular, single-phase ionisation modelling does not accurately capture the underlying {\it distribution\/} of metallicities and gas physical conditions \citep[see][]{haislmaier21, Sameer21}, even if it might capture the mean values \citep{marra21b, Sameer21}. We expect that more complex, multi-phase modelling approaches \citep[e.g.,][]{Zahedy19, Zahedy21, haislmaier21, Sameer21, Sameer22, nielsen22}, will provide an improved characterisation of CGM gas. Specifically, the ionisation modelling method being developed by \citet{Sameer21, Sameer22} allows for multiple clouds in different phases of gas to contribute to a single VP component.  Their treatment is the most consistent with the line formation processes for individual absorption components as seen in the simulations.

Even so, the multi-cloud multi-phase method of \citet{Sameer21, Sameer22} implements a minimalist Bayesian approach to the number of clouds contributing to an absorption component, whereas the simulations suggest that the number can range from one to several spatially isolated clouds. As suggested by the simulations, the line formation process is complex and involves multiple spatially separated clouds; the absorption line data simply do not provide the level of information required to fully reveal the underlying true gas structures. It is likely that no matter how high the signal-to-noise ratio in the highest-resolution spectra currently available with 10 or 30-meter class telescopes, there will still be hidden information on the complexity of the spatial distribution of the gas as illustrated in Figure~\ref{spec_diff_comps}. 

The line formation physics is further complicated by the fact that each of the multiple spatially segregated clouds  collectively giving rise to an individual absorption component exhibits a range of physical properties. Consider the example {\SiII}~$\lambda 1260$, {\CIV}~$\lambda 1548$, and {\OVI}~$\lambda 1031$ absorption profiles and their absorption component spatially distinct clouds shown in Figure~\ref{spec_abs}. We again present these profiles in Figure~\ref{AbsCellProps}, where we also show the distribution of the physical properties of the absorbing gas as a function of LOS velocity. From top to bottom, these quantities are the absorbing gas cell hydrogen number density, $n_{\tH}$, temperature, $T$, metallicity, $Z/Z_{\odot}$, and LOS position, $\Delta S$. In all panels, the data corresponding to each absorption component are colour shaded and separated by red vertical lines. For a given absorption component, the point types designate to which spatially isolated cloud the cells belong.

The important information in Figure~\ref{AbsCellProps} is that LOS velocity aligned gas giving rise to an absorption component can have a range of kinematic, chemical, and gas-phase conditions. For {\OVI}, we can see that there is a large spread of temperature values in each absorption component. Consider the absorption component centred at $v_{\hbox{\tiny LOS}} \simeq 35$~{\kms} of the  {\OVI}~$\lambda 1031$ absorption profile. The two contributing clouds, separated by roughly $\Delta S \simeq 10$~kpc, have similar metallicities, but their $n_{\tH}$ differ by an order of magnitude. Furthermore, {\it both\/} clouds have a bimodality in their temperatures. Similar complexities can be seen for the absorption components of all three ions, though the low-ionisation clouds giving rise to {\SiII} absorption exhibit the least dispersion in their physical conditions.  The gas probed by {\CIV} exhibits the most complexity and variation in its physical conditions. 

It is also interesting to point out the density and metallicity gradients across the full {\OVI}~$\lambda 1031$ absorption profile, where $\log n_{\tH}$ ranges from $-2.5$ to $-4.75$ [cm$^{-3}$] and $\log Z/Z_{\odot}$ ranges from $-3.25$ to $-1.6$ across the LOS velocity range $v_{\hbox{\tiny LOS}}  \simeq -35$ to $+45$~{\kms} and LOS position $\Delta S \simeq +10$ to $-30$~kpc relative to the plane of the sky. Though further investigation of the distributions of $n_{\tH}$, $T$, and $Z/Z_{\odot}$ in the clouds giving rise to an absorption component is beyond the scope of this work, it would be of interest in future work to deepen our characterisation of the absorbing gas for improving our understanding of the formation of absorption lines in the CGM. 

Finally, we discuss the common mass PDF shown in Figure~\ref{colocated}. This function is designed to provide insights into the multi-phase nature of the individual absorption components of different ions having LOS velocity alignments. When the percent common mass, $f_{\hbox{\tiny CM}}$, is 95--100\%, this would constitute single-phase gas, whereas the smaller the value of $f_{\hbox{\tiny CM}}$, the greater the degree of multi-phase ionisation conditions. 

LOS velocity alignment of absorption components does not imply spatially coincident absorption clouds. The question is, to what degree do LOS velocity aligned absorption components of lower- and higher-ionisation ions arise in the same parcel of gas, i.e., are a result of the ionisation conditions in a single-phase cloud?  Or, alternately, to what degree do they arise in (i) physically separated clouds, or (ii) portions of clouds separated by ionisation fronts? Both scenarios represent multi-phase ionisation conditions. The latter refers to clouds with ionisation structure such that the lower and higher-ionisation species are physically segregated but reside within a single cloud (as we have defined them in this work).

Naive expectations are that the smaller (greater) the difference in the ionisation potential of the ions, the more (less) likely they reside co-spatially. Quantifying this co-spatial overlap in terms of the gas mass, we find this expectation to hold. As seen in Figure~\ref{colocated}, there is less than a 1\% chance that the gas mass in common to LOS velocity aligned absorption components of {\SiII}~$\lambda 1260$ and {\OVI}~$\lambda 1031$ profiles is as high as 40\%.  In fact, for roughly 99\% of absorption components, less than 5\% of the gas mass is in common for these ions.  This indicates that multi-phase ionisation conditions apply to {\SiII} and {\OVI} components that align in LOS velocity.

For the low-ionisation Si$^{+}$ and intermediate-ionisation C$^{+3}$ ions, as observed in {\SiII}~$\lambda 1260$ and {\CIV}~$\lambda 1548$ absorption, the common mass PDF is relatively flat in that  LOS velocity aligned absorption components have a roughly equal probability of about 2\% of sharing anywhere from 5--95\% of their absorbing gas mass. However, the highest probabilities are that either 0--5\% of their mass is in common or 95-100\% of their mass is in common.  This PDF informs us that these two ions exhibit a range of single-phase and multi-phase ionisation conditions, with a slight preference (45\%) of a partial multi-phase condition (shared mass of 5--95\%), but with a tendency toward pure multi-phase conditions (40\%) or pure single-phase conditions (15\%).

A similar common mass PDF is found for the high-ionisation O$^{+5}$ and intermediate-ionisation C$^{+3}$ ions. The common mass PDF is flat with the strongest peak probability at 0--5\% and a second peak at 95-100\% of their mass in common. For these ions, 70\% of the LOS velocity aligned absorption components have between 0--5\% of the mass gas in common, the condition that would classify as multi-phase absorption. Only 4\% have between 95-100\% of their mass in common, the condition that would characterise as single-phase conditions.

\subsection{Caveats}

Endemic to all studies of hydrodynamic simulations are the limitations of a finite resolution and both the prioritisation and approximations inherent in the implementation of sub-grid physics. We fully appreciate that the exact manifestations of the distributions of the number of clouds per absorption component, the values of the statistical descriptors characterising the TPSF, and the shapes of the common mass PDFs may have a degree of specificity to the simulations we have adopted for this study. For example, we have used $z=1$ low-mass dwarf galaxies with current low star-formation rates. The feedback recipes and  the resolution likely have some governing control over the development of the CGM surrounding these galaxies. For example, there are no shock fronts in the CGM of these galaxies, nor are there large mass concentrations of metal-enriched collisionally ionised gas. Both of these features are expected to become more prominent in the CGM at lower redshifts and/or higher galaxy masses \citep[e.g.,][]{birnboim03, Dekel06, keres05, keres09, oppenheimer10, Oppenheimer16, Nelson2018, peroux20}.  Too low of a resolution or too aggressive of a gas heating function, for example, can suppress the condensation of cool gas structures or smooth the spatial variations in the gas density, temperature, and chemical enrichment \citep[e.g.,][]{ceverino09, Hummels19, peeples19, vandevoort19, nelson20}. With regards to the conclusions we have drawn in this work, it is possible that the details of our results have resolution dependency, sub-grid physics dependency, and even galaxy property (mass, star formation rate, etc) dependency. 

With due respect to the complications and remaining unknowns as to resolution and sub-grid physics effects on the realism of the simulated CGM, we are confident that the essence of our work holds. With regards to the definition of absorbing clouds, we have adopted an objective method that accounts for the variable resolution and that segregates the gas structures in the context of the gas cell sizes. To our knowledge, no other works have investigated the relationship between absorption lines and absorbing gas in the level of detail we have presented here, meaning a direct and objective examination of the correspondence between individual absorption components and individual absorbing clouds along a line of sight through the CGM.

The fact that the details may have dependency on aforementioned factors highlights the fact that there is much work remaining in order to improve our understanding and characterisation of the CGM through simulations using absorption lines. Our goal has been to forge a methodology on how to proceed, which is to say, to begin the process of dissecting and characterising the physical conditions of the absorbing gas associated with individual absorption components.  As the physics of the CGM is better understood, and simulations are improved to better reflect the true nature of the CGM, we can then begin to explore questions such as: what signatures can be identified in the absorption line data that allow direct insights into the complex velocity fields and spatially discrete structures that map to a given LOS velocity?  What are plausible new approaches to absorption line analysis that incorporate a more realistic view of the CGM?  What are the implications for our understanding of the CGM and the baryon cycle once such signatures can be employed?

\section{Summary and Conclusions}
\label{Conclusion}

Our aim with this study was to advance our appreciation and insight into the efficacy of the techniques applied to quasar absorption lines for determining the physical conditions of the absorbing gas in the CGM. In particular, we confronted the assumption that an individual absorption component in a complex absorption profile maps to a single, spatially isolated, cloud-like structure by emulating the quasar absorption line technique in cosmological simulations. Synthetic ``COS'' absorption line spectra with ${\it SNR} = 30$ were generated through the CGM  \citep[see][]{churchill15, rachel_thesis} of two simulated dwarf galaxies in the high-resolution hydrodynamic cosmological simulations of \citet{Trujillo15}. Objective methods \citep[see][]{schneider93, weakI} were implemented to identify and quantify absorption lines in the synthetic spectra. A $5\sigma$ minimum rest-frame equivalent width detection threshold of 0.05~{\AA} was applied to define our sample.

As described in Section~\ref{DefAbsCells}, we objectively identified the absorbing gas cells pierced by the LOS that contribute to the {\it detected\/} absorption profiles \citep{churchill15, rachel_thesis}. As observed absorption lines represent only the gas giving rise to the actually detected absorption in the spectra, it seems the theoretical studies of simulations should adopt this physical fact. 
Our study focused on the {\SiII}~$\lambda 1260$ and {\CIVdblt},  {\OVIdblt} absorption profiles. We developed and applied an objective methodology to define absorption components in the absorption profiles, allowing us to identify and characterised the spatial and physical properties of the absorbing cells associated with each absorption component based on their LOS velocity. Our method flexibly accounts for the varying resolution of the gas cells in the zoom-in simulations. Overall, we studied a total of 1,302 LOS containing a total of 7,755 absorption components.

Our results and conclusions can be summarised as follows: 

(1) For a given absorption component, we found that one to several spatially distinct cloud-like structures give rise to the absorption component. The majority (52\%) of {\SiII} absorption components arise from a single, spatially isolated cloud, whereas the majority of absorption components of both {\CIV} and {\OVI} (68\% and 72\%, respectively) arise from two or more spatially distinct clouds, with two being the most frequent number (28\% and 33\% of absorption components, respectively). These results are presented in Figure~\ref{numcomps} and Table~\ref{tab:pmf}.  The important finding here is that a given absorption component does not necessarily map to a single spatially isolated gas cloud; it may map to several spatially distinct gas clouds, with potentially different gas properties, along the LOS that happen to be aligned in LOS velocity.

(2) For absorption components arising from multiple spatially distinct cloud-like structures, the mode of the distribution of cloud-cloud LOS separation distances between their density weighted centres (see Eq.~\ref{eq:DScenter}) is 1.3, 5, and 6 kpc for {\SiII}, {\CIV}, and {\OVI} absorption, respectively. For {\SiII} absorption components, 50\% of the multiple cloud-cloud LOS separations lie within 3~kpc and 90\% lie within 7~kpc. For {\CIV} and {\OVI} absorption components, 50\% lie within 9 and 12~kpc, respectively, and 90\% lie within 21 and 32~kpc, respectively. The maximum LOS separations are on the order of $0.4R_{vir}$, $1R_{vir}$, and $2.3R_{vir}$, for {\SiII}, {\CIV}, and {\OVI}, respectively.  These results are presented in Figure~\ref{freq_hist} and Table~\ref{tab:tpsf}. It is possible the exact shape of the cloud-cloud LOS separation distribution measured in this work is dependent on the properties of the simulated CGM and therefore on the simulated galaxies we have studied. However, we expect that the general trends are a universal feature of CGM gas that may scale with simulate galaxy properties. Characterising this anticipated scaling is a future investigation we aim to pursue.

(3) For absorption components from different ions that aligned in LOS velocity, we examined how much mass in their absorbing clouds was in common (see Eq.~\ref{eq:fCM}). We then computed a ``common mass probability distribution function,'' which measures the probability that two ions with a LOS velocity aligned absorption component will have a given percentage of absorbing gas mass in common. We find less than 1\% of {\SiII} and {\OVI} LOS velocity aligned absorption components have 50\% or more of their gas mass in common.  For {\SiII} and {\CIV}, roughly one-third (34\%) have 50\% or more of their gas mass in common.  And for  {\OVI} and {\CIV}, roughly 15\% have 50\% or more of their gas mass in common.  These results, presented in Figure~\ref{colocated} and Table~\ref{tab:tpsf}, provide quantifiable insights into the relative number of velocity aligned components for different ions arising in a single gas phase or arise in multi-phase gas.  The greater the percentage of components that have higher fractions of common mass, the more confidently they can be approximated as single-phase absorption systems for the purposes of ionisation modelling.

Future work will reveal how sensitive our results may be to resolution, sub-grid physics, and galaxy properties. Similar analysis using different simulations would advance the conversation on these important issues.  In the meantime, our work indicates that there is tension between the assumptions applied to the majority of absorption line analyses, namely that a single absorption component maps to a single spatially distinct gas cloud.  Furthermore, the tension increases when one considers that the spatially distinct gas clouds are assumed to be single valued in their density, temperature, metallicity, and ionisation condition. Our results clearly show that this is an incorrect view of the line formation process, as the absorbing gas exhibits a range of physical properties across one to several discrete gas structures. Additional work is required to better characterise this tension and to ultimately apply lessons from insights gained using hydrodynamic simulations to observational analyses techniques and studies of the baryon cycle governing the evolution of galaxies.

\section*{Acknowledgements}
RM holds an American Fellowship from AAUW. This material is based upon work supported by the National Science Foundation under Grant No.\ 1517816 issued to CWC. GGK and NMN acknowledge the support of the Australian Research Council Centre of Excellence for All Sky Astrophysics in 3 Dimensions (ASTRO 3D), through project number CE170100013. STG gratefully acknowledges funding from the European Research Council (ERC) under the European Union’s Horizon 2020 research and innovation programme via the ERC Starting Grant MUSTANG (grant agreement number 714907).

\section*{Data Availability}
The data utilised in this paper will be shared on reasonable request
to the corresponding author.

\bibliographystyle{mnras}
\bibliography{refs} 

\begin{thebibliography}{}
\makeatletter
\relax
\def\mn@urlcharsother{\let\do\@makeother \do\$\do\&\do\#\do\^\do\_\do\%\do\~}
\def\mn@doi{\begingroup\mn@urlcharsother \@ifnextchar [ {\mn@doi@}
  {\mn@doi@[]}}
\def\mn@doi@[#1]#2{\def\@tempa{#1}\ifx\@tempa\@empty \href
  {http://dx.doi.org/#2} {doi:#2}\else \href {http://dx.doi.org/#2} {#1}\fi
  \endgroup}
\def\mn@eprint#1#2{\mn@eprint@#1:#2::\@nil}
\def\mn@eprint@arXiv#1{\href {http://arxiv.org/abs/#1} {{\tt arXiv:#1}}}
\def\mn@eprint@dblp#1{\href {http://dblp.uni-trier.de/rec/bibtex/#1.xml}
  {dblp:#1}}
\def\mn@eprint@#1:#2:#3:#4\@nil{\def\@tempa {#1}\def\@tempb {#2}\def\@tempc
  {#3}\ifx \@tempc \@empty \let \@tempc \@tempb \let \@tempb \@tempa \fi \ifx
  \@tempb \@empty \def\@tempb {arXiv}\fi \@ifundefined
  {mn@eprint@\@tempb}{\@tempb:\@tempc}{\expandafter \expandafter \csname
  mn@eprint@\@tempb\endcsname \expandafter{\@tempc}}}

\bibitem[\protect\citeauthoryear{{Asplund}, {Grevesse}, {Sauval}  \&
  {Scott}}{{Asplund} et~al.}{2009}]{Asplund09}
{Asplund} M.,  {Grevesse} N.,  {Sauval} A.~J.,   {Scott} P.,  2009, \mn@doi
  [\araa] {10.1146/annurev.astro.46.060407.145222}, \href
  {http://adsabs.harvard.edu/abs/2009ARA%26A..47..481A} {47, 481}

\bibitem[\protect\citeauthoryear{{Bainbridge} \& {Webb}}{{Bainbridge} \&
  {Webb}}{2017}]{bainbridge17}
{Bainbridge} M.~B.,  {Webb} J.~K.,  2017, \mn@doi [\mnras]
  {10.1093/mnras/stx179}, \href
  {https://ui.adsabs.harvard.edu/abs/2017MNRAS.468.1639B} {468, 1639}

\bibitem[\protect\citeauthoryear{{Bergeron} \& {Stasi{\'n}ska}}{{Bergeron} \&
  {Stasi{\'n}ska}}{1986}]{bergeron86}
{Bergeron} J.,  {Stasi{\'n}ska} G.,  1986, \aap, \href
  {https://ui.adsabs.harvard.edu/abs/1986A&A...169....1B} {169, 1}

\bibitem[\protect\citeauthoryear{{Birnboim} \& {Dekel}}{{Birnboim} \&
  {Dekel}}{2003}]{birnboim03}
{Birnboim} Y.,  {Dekel} A.,  2003, \mn@doi [\mnras]
  {10.1046/j.1365-8711.2003.06955.x}, \href
  {http://adsabs.harvard.edu/abs/2003MNRAS.345..349B} {345, 349}

\bibitem[\protect\citeauthoryear{{Bordoloi} et~al.,}{{Bordoloi}
  et~al.}{2014}]{bordoloi-cosdwarfs}
{Bordoloi} R.,  et~al., 2014, \mn@doi [\apj] {10.1088/0004-637X/796/2/136},
  \href {http://adsabs.harvard.edu/abs/2014ApJ...796..136B} {796, 136}

\bibitem[\protect\citeauthoryear{{Carswell} \& {Webb}}{{Carswell} \&
  {Webb}}{2014}]{vpfit}
{Carswell} R.~F.,  {Webb} J.~K.,  2014, {VPFIT: Voigt profile fitting program},
  Astrophysics Source Code Library (\mn@eprint {ascl} {1408.015})

\bibitem[\protect\citeauthoryear{{Carswell}, {Lanzetta}, {Parnell}  \&
  {Webb}}{{Carswell} et~al.}{1991}]{carswell91}
{Carswell} R.~F.,  {Lanzetta} K.~M.,  {Parnell} H.~C.,   {Webb} J.~K.,  1991,
  \mn@doi [\apj] {10.1086/169868}, \href
  {https://ui.adsabs.harvard.edu/abs/1991ApJ...371...36C} {371, 36}

\bibitem[\protect\citeauthoryear{{Ceverino} \& {Klypin}}{{Ceverino} \&
  {Klypin}}{2009}]{ceverino09}
{Ceverino} D.,  {Klypin} A.,  2009, \mn@doi [\apj]
  {10.1088/0004-637X/695/1/292}, \href
  {http://adsabs.harvard.edu/abs/2009ApJ...695..292C} {695, 292}

\bibitem[\protect\citeauthoryear{{Ceverino}, {Dekel}  \& {Bournaud}}{{Ceverino}
  et~al.}{2010}]{Ceverino10}
{Ceverino} D.,  {Dekel} A.,   {Bournaud} F.,  2010, \mn@doi [\mnras]
  {10.1111/j.1365-2966.2010.16433.x}, \href
  {http://adsabs.harvard.edu/abs/2010MNRAS.404.2151C} {404, 2151}

\bibitem[\protect\citeauthoryear{{Ceverino}, {Klypin}, {Klimek},
  {Trujillo-Gomez}, {Churchill}, {Primack}  \& {Dekel}}{{Ceverino}
  et~al.}{2014}]{ceverino14}
{Ceverino} D.,  {Klypin} A.,  {Klimek} E.~S.,  {Trujillo-Gomez} S.,
  {Churchill} C.~W.,  {Primack} J.,   {Dekel} A.,  2014, \mn@doi [\mnras]
  {10.1093/mnras/stu956}, \href
  {http://adsabs.harvard.edu/abs/2014MNRAS.442.1545C} {442, 1545}

\bibitem[\protect\citeauthoryear{{Charlton}, {Mellon}, {Rigby}  \&
  {Churchill}}{{Charlton} et~al.}{2000}]{charlton00}
{Charlton} J.~C.,  {Mellon} R.~R.,  {Rigby} J.~R.,   {Churchill} C.~W.,  2000,
  \mn@doi [\apj] {10.1086/317825}, \href
  {https://ui.adsabs.harvard.edu/abs/2000ApJ...545..635C} {545, 635}

\bibitem[\protect\citeauthoryear{{Charlton}, {Ding}, {Zonak}, {Churchill},
  {Bond}  \& {Rigby}}{{Charlton} et~al.}{2003}]{charlton03}
{Charlton} J.~C.,  {Ding} J.,  {Zonak} S.~G.,  {Churchill} C.~W.,  {Bond}
  N.~A.,   {Rigby} J.~R.,  2003, \mn@doi [\apj] {10.1086/374353}, \href
  {https://ui.adsabs.harvard.edu/abs/2003ApJ...589..111C} {589, 111}

\bibitem[\protect\citeauthoryear{{Churchill}}{{Churchill}}{1997}]{cwc-thesis}
{Churchill} C.~W.,  1997, PhD thesis, {University of California, Santa Cruz}

\bibitem[\protect\citeauthoryear{Churchill \& Charlton}{Churchill \&
  Charlton}{1999}]{cwc_charlton}
Churchill C.~W.,  Charlton J.~C.,  1999, The Astronomical Journal, 118, 59

\bibitem[\protect\citeauthoryear{{Churchill} \& {Vogt}}{{Churchill} \&
  {Vogt}}{2001}]{cv01}
{Churchill} C.~W.,  {Vogt} S.~S.,  2001, \mn@doi [\aj] {10.1086/321174}, \href
  {http://adsabs.harvard.edu/abs/2001AJ....122..679C} {122, 679}

\bibitem[\protect\citeauthoryear{{Churchill}, {Rigby}, {Charlton}  \&
  {Vogt}}{{Churchill} et~al.}{1999}]{weakI}
{Churchill} C.~W.,  {Rigby} J.~R.,  {Charlton} J.~C.,   {Vogt} S.~S.,  1999,
  \mn@doi [\apjs] {10.1086/313168}, \href
  {http://adsabs.harvard.edu/abs/1999ApJS..120...51C} {120, 51}

\bibitem[\protect\citeauthoryear{{Churchill}, {Mellon}, {Charlton}, {Jannuzi},
  {Kirhakos}, {Steidel}  \& {Schneider}}{{Churchill}
  et~al.}{2000}]{churchill00}
{Churchill} C.~W.,  {Mellon} R.~R.,  {Charlton} J.~C.,  {Jannuzi} B.~T.,
  {Kirhakos} S.,  {Steidel} C.~C.,   {Schneider} D.~P.,  2000, \mn@doi [\apj]
  {10.1086/317120}, \href {http://adsabs.harvard.edu/abs/2000ApJ...543..577C}
  {543, 577}

\bibitem[\protect\citeauthoryear{{Churchill}, {Vogt}  \&
  {Charlton}}{{Churchill} et~al.}{2003a}]{cvc03}
{Churchill} C.~W.,  {Vogt} S.~S.,   {Charlton} J.~C.,  2003a, \mn@doi [\aj]
  {10.1086/345513}, \href {http://adsabs.harvard.edu/abs/2003AJ....125...98C}
  {125, 98}

\bibitem[\protect\citeauthoryear{{Churchill}, {Mellon}, {Charlton}  \&
  {Vogt}}{{Churchill} et~al.}{2003b}]{churchill03}
{Churchill} C.~W.,  {Mellon} R.~R.,  {Charlton} J.~C.,   {Vogt} S.~S.,  2003b,
  \mn@doi [\apj] {10.1086/376444}, \href
  {https://ui.adsabs.harvard.edu/abs/2003ApJ...593..203C} {593, 203}

\bibitem[\protect\citeauthoryear{{Churchill}, {Kacprzak}, {Steidel}, {Spitler},
  {Holtzman}, {Nielsen}  \& {Trujillo-Gomez}}{{Churchill}
  et~al.}{2012}]{cwc1317b}
{Churchill} C.~W.,  {Kacprzak} G.~G.,  {Steidel} C.~C.,  {Spitler} L.~R.,
  {Holtzman} J.,  {Nielsen} N.~M.,   {Trujillo-Gomez} S.,  2012, \mn@doi [\apj]
  {10.1088/0004-637X/760/1/68}, \href
  {http://adsabs.harvard.edu/abs/2012ApJ...760...68C} {760, 68}

\bibitem[\protect\citeauthoryear{{Churchill}, {Klimek}, {Medina}  \& {Vander
  Vliet}}{{Churchill} et~al.}{2014}]{cwc14}
{Churchill} C.~W.,  {Klimek} E.,  {Medina} A.,   {Vander Vliet} J.~R.,  2014,
  preprint, \href {http://adsabs.harvard.edu/abs/2014arXiv1409.0916C} {}
  (\mn@eprint {arXiv} {1409.0916})

\bibitem[\protect\citeauthoryear{{Churchill}, {Vander Vliet}, {Trujillo-Gomez},
  {Kacprzak}  \& {Klypin}}{{Churchill} et~al.}{2015}]{churchill15}
{Churchill} C.~W.,  {Vander Vliet} J.~R.,  {Trujillo-Gomez} S.,  {Kacprzak}
  G.~G.,   {Klypin} A.,  2015, \mn@doi [\apj] {10.1088/0004-637X/802/1/10},
  \href {http://adsabs.harvard.edu/abs/2015ApJ...802...10C} {802, 10}

\bibitem[\protect\citeauthoryear{{Churchill}, {Evans}, {Stemock}, {Nielsen},
  {Kacprzak}  \& {Murphy}}{{Churchill} et~al.}{2020}]{churchill20}
{Churchill} C.~W.,  {Evans} J.~L.,  {Stemock} B.,  {Nielsen} N.~M.,  {Kacprzak}
  G.~G.,   {Murphy} M.~T.,  2020, \mn@doi [\apj] {10.3847/1538-4357/abbb34},
  \href {https://ui.adsabs.harvard.edu/abs/2020ApJ...904...28C} {904, 28}

\bibitem[\protect\citeauthoryear{{Crighton}, {Hennawi}  \&
  {Prochaska}}{{Crighton} et~al.}{2013}]{crighton13}
{Crighton} N. H.~M.,  {Hennawi} J.~F.,   {Prochaska} J.~X.,  2013, \mn@doi
  [\apjl] {10.1088/2041-8205/776/2/L18}, \href
  {https://ui.adsabs.harvard.edu/abs/2013ApJ...776L..18C} {776, L18}

\bibitem[\protect\citeauthoryear{{Davis}, {Jiang}, {Stone}  \&
  {Murray}}{{Davis} et~al.}{2014}]{Davis14}
{Davis} S.~W.,  {Jiang} Y.-F.,  {Stone} J.~M.,   {Murray} N.,  2014, \mn@doi
  [\apj] {10.1088/0004-637X/796/2/107}, \href
  {https://ui.adsabs.harvard.edu/abs/2014ApJ...796..107D} {796, 107}

\bibitem[\protect\citeauthoryear{{Dekel} \& {Birnboim}}{{Dekel} \&
  {Birnboim}}{2006}]{Dekel06}
{Dekel} A.,  {Birnboim} Y.,  2006, \mn@doi [\mnras]
  {10.1111/j.1365-2966.2006.10145.x}, \href
  {https://ui.adsabs.harvard.edu/abs/2006MNRAS.368....2D} {368, 2}

\bibitem[\protect\citeauthoryear{{Ding}, {Charlton}, {Bond}, {Zonak}  \&
  {Churchill}}{{Ding} et~al.}{2003a}]{ding03a}
{Ding} J.,  {Charlton} J.~C.,  {Bond} N.~A.,  {Zonak} S.~G.,   {Churchill}
  C.~W.,  2003a, \mn@doi [\apj] {10.1086/368250}, \href
  {https://ui.adsabs.harvard.edu/abs/2003ApJ...587..551D} {587, 551}

\bibitem[\protect\citeauthoryear{{Ding}, {Charlton}, {Churchill}  \&
  {Palma}}{{Ding} et~al.}{2003b}]{ding03b}
{Ding} J.,  {Charlton} J.~C.,  {Churchill} C.~W.,   {Palma} C.,  2003b, \mn@doi
  [\apj] {10.1086/375028}, \href
  {https://ui.adsabs.harvard.edu/abs/2003ApJ...590..746D} {590, 746}

\bibitem[\protect\citeauthoryear{{Ding}, {Charlton}  \& {Churchill}}{{Ding}
  et~al.}{2005}]{ding05}
{Ding} J.,  {Charlton} J.~C.,   {Churchill} C.~W.,  2005, \mn@doi [\apj]
  {10.1086/427623}, \href
  {https://ui.adsabs.harvard.edu/abs/2005ApJ...621..615D} {621, 615}

\bibitem[\protect\citeauthoryear{{Draine}}{{Draine}}{2011}]{Draine}
{Draine} B.~T.,  2011, Physics of the Interstellar and Intergalactic Medium.
Princeton University Press

\bibitem[\protect\citeauthoryear{{Ferland}, {Korista}, {Verner}, {Ferguson},
  {Kingdon}  \& {Verner}}{{Ferland} et~al.}{1998}]{Ferland98}
{Ferland} G.~J.,  {Korista} K.~T.,  {Verner} D.~A.,  {Ferguson} J.~W.,
  {Kingdon} J.~B.,   {Verner} E.~M.,  1998, \mn@doi [\pasp] {10.1086/316190},
  \href {http://adsabs.harvard.edu/abs/1998PASP..110..761F} {110, 761}

\bibitem[\protect\citeauthoryear{{Ferland} et~al.,}{{Ferland}
  et~al.}{2013}]{Ferland13}
{Ferland} G.~J.,  et~al., 2013, Revista Mexicana de Astronomia y Astrofisica,
  \href {http://adsabs.harvard.edu/abs/2013RMxAA..49..137F} {49, 137}

\bibitem[\protect\citeauthoryear{{Ferland} et~al.,}{{Ferland}
  et~al.}{2017}]{Ferland17}
{Ferland} G.~J.,  et~al., 2017, \rmxaa, \href
  {https://ui.adsabs.harvard.edu/abs/2017RMxAA..53..385F} {53, 385}

\bibitem[\protect\citeauthoryear{{Fontana} \& {Ballester}}{{Fontana} \&
  {Ballester}}{1995}]{fontana95}
{Fontana} A.,  {Ballester} P.,  1995, The Messenger, \href
  {https://ui.adsabs.harvard.edu/abs/1995Msngr..80...37F} {80, 37}

\bibitem[\protect\citeauthoryear{{Fox}, {Wakker}, {Savage}, {Tripp}, {Sembach}
  \& {Bland-Hawthorn}}{{Fox} et~al.}{2005}]{Fox05}
{Fox} A.~J.,  {Wakker} B.~P.,  {Savage} B.~D.,  {Tripp} T.~M.,  {Sembach}
  K.~R.,   {Bland-Hawthorn} J.,  2005, \mn@doi [\apj] {10.1086/431915}, \href
  {https://ui.adsabs.harvard.edu/abs/2005ApJ...630..332F} {630, 332}

\bibitem[\protect\citeauthoryear{{Fox} et~al.,}{{Fox} et~al.}{2015}]{fox15}
{Fox} A.~J.,  et~al., 2015, \mn@doi [\apjl] {10.1088/2041-8205/799/1/L7}, \href
  {http://adsabs.harvard.edu/abs/2015ApJ...799L...7F} {799, L7}

\bibitem[\protect\citeauthoryear{{Fumagalli}, {O'Meara}  \&
  {Prochaska}}{{Fumagalli} et~al.}{2016}]{fumagalli16}
{Fumagalli} M.,  {O'Meara} J.~M.,   {Prochaska} J.~X.,  2016, \mn@doi [\mnras]
  {10.1093/mnras/stv2616}, \href
  {https://ui.adsabs.harvard.edu/abs/2016MNRAS.455.4100F} {455, 4100}

\bibitem[\protect\citeauthoryear{{Gaikwad}, {Srianand}, {Choudhury}  \&
  {Khaire}}{{Gaikwad} et~al.}{2017}]{gaikwad17}
{Gaikwad} P.,  {Srianand} R.,  {Choudhury} T.~R.,   {Khaire} V.,  2017, \mn@doi
  [\mnras] {10.1093/mnras/stx248}, \href
  {https://ui.adsabs.harvard.edu/abs/2017MNRAS.467.3172G} {467, 3172}

\bibitem[\protect\citeauthoryear{{Glidden}, {Cooper}, {Cooksey}, {Simcoe}  \&
  {O'Meara}}{{Glidden} et~al.}{2016}]{glidden16}
{Glidden} A.,  {Cooper} T.~J.,  {Cooksey} K.~L.,  {Simcoe} R.~A.,   {O'Meara}
  J.~M.,  2016, \mn@doi [\apj] {10.3847/1538-4357/833/2/270}, \href
  {https://ui.adsabs.harvard.edu/abs/2016ApJ...833..270G} {833, 270}

\bibitem[\protect\citeauthoryear{{Gnat} \& {Sternberg}}{{Gnat} \&
  {Sternberg}}{2007}]{GnatSternberg07}
{Gnat} O.,  {Sternberg} A.,  2007, \mn@doi [\apjs] {10.1086/509786}, \href
  {https://ui.adsabs.harvard.edu/abs/2007ApJS..168..213G} {168, 213}

\bibitem[\protect\citeauthoryear{{Green} et~al.,}{{Green}
  et~al.}{2012}]{green-cos}
{Green} J.~C.,  et~al., 2012, \mn@doi [\apj] {10.1088/0004-637X/744/1/60},
  \href {https://ui.adsabs.harvard.edu/abs/2012ApJ...744...60G} {744, 60}

\bibitem[\protect\citeauthoryear{{Groves}, {Dopita}, {Sutherland}, {Kewley},
  {Fischera}, {Leitherer}, {Brandl}  \& {van Breugel}}{{Groves}
  et~al.}{2008}]{Groves08}
{Groves} B.,  {Dopita} M.~A.,  {Sutherland} R.~S.,  {Kewley} L.~J.,  {Fischera}
  J.,  {Leitherer} C.,  {Brandl} B.,   {van Breugel} W.,  2008, \mn@doi [\apjs]
  {10.1086/528711}, \href
  {https://ui.adsabs.harvard.edu/abs/2008ApJS..176..438G} {176, 438}

\bibitem[\protect\citeauthoryear{{Haardt} \& {Madau}}{{Haardt} \&
  {Madau}}{2005}]{HaardtMadau2005}
{Haardt} F.,  {Madau} P.,  2005, unpublished spectra in 2005 August update to
  Haardt \& Madau (2001) and included in the photoionization code CLOUDY

\bibitem[\protect\citeauthoryear{{Haislmaier}, {Tripp}, {Katz}, {Prochaska},
  {Burchett}, {O'Meara}  \& {Werk}}{{Haislmaier} et~al.}{2021}]{haislmaier21}
{Haislmaier} K.~J.,  {Tripp} T.~M.,  {Katz} N.,  {Prochaska} J.~X.,  {Burchett}
  J.~N.,  {O'Meara} J.~M.,   {Werk} J.~K.,  2021, \mn@doi [\mnras]
  {10.1093/mnras/staa3544}, \href
  {https://ui.adsabs.harvard.edu/abs/2021MNRAS.502.4993H} {502, 4993}

\bibitem[\protect\citeauthoryear{{Howarth}}{{Howarth}}{2015}]{howarth15}
{Howarth} I.~D.,  2015, {VAPID: Voigt Absorption-Profile [Interstellar]
  Dabbler} (\mn@eprint {ascl} {1506.010})

\bibitem[\protect\citeauthoryear{{Hummels} et~al.,}{{Hummels}
  et~al.}{2019}]{Hummels19}
{Hummels} C.~B.,  et~al., 2019, \mn@doi [\apj] {10.3847/1538-4357/ab378f},
  \href {https://ui.adsabs.harvard.edu/abs/2019ApJ...882..156H} {882, 156}

\bibitem[\protect\citeauthoryear{{Kacprzak}, {Churchill}, {Ceverino},
  {Steidel}, {Klypin}  \& {Murphy}}{{Kacprzak} et~al.}{2010}]{kacprzak10}
{Kacprzak} G.~G.,  {Churchill} C.~W.,  {Ceverino} D.,  {Steidel} C.~C.,
  {Klypin} A.,   {Murphy} M.~T.,  2010, \mn@doi [\apj]
  {10.1088/0004-637X/711/2/533}, \href
  {https://ui.adsabs.harvard.edu/abs/2010ApJ...711..533K} {711, 533}

\bibitem[\protect\citeauthoryear{{Kacprzak}, {Churchill}, {Steidel}, {Spitler}
  \& {Holtzman}}{{Kacprzak} et~al.}{2012a}]{Kacprzak12a}
{Kacprzak} G.~G.,  {Churchill} C.~W.,  {Steidel} C.~C.,  {Spitler} L.~R.,
  {Holtzman} J.~A.,  2012a, \mn@doi [\mnras]
  {10.1111/j.1365-2966.2012.21945.x}, \href
  {https://ui.adsabs.harvard.edu/abs/2012MNRAS.427.3029K} {427, 3029}

\bibitem[\protect\citeauthoryear{{Kacprzak}, {Churchill}  \&
  {Nielsen}}{{Kacprzak} et~al.}{2012b}]{kcn12}
{Kacprzak} G.~G.,  {Churchill} C.~W.,   {Nielsen} N.~M.,  2012b, \mn@doi
  [\apjl] {10.1088/2041-8205/760/1/L7}, \href
  {http://adsabs.harvard.edu/abs/2012ApJ...760L...7K} {760, L7}

\bibitem[\protect\citeauthoryear{Kacprzak et~al.,}{Kacprzak
  et~al.}{2019}]{Kacprzak_2019}
Kacprzak G.~G.,  et~al., 2019, \mn@doi [The Astrophysical Journal]
  {10.3847/1538-4357/aaf1a6}, 870, 137

\bibitem[\protect\citeauthoryear{{Kere{\v s}}, {Katz}, {Weinberg}  \&
  {Dav{\'e}}}{{Kere{\v s}} et~al.}{2005}]{keres05}
{Kere{\v s}} D.,  {Katz} N.,  {Weinberg} D.~H.,   {Dav{\'e}} R.,  2005, \mn@doi
  [\mnras] {10.1111/j.1365-2966.2005.09451.x}, \href
  {http://adsabs.harvard.edu/abs/2005MNRAS.363....2K} {363, 2}

\bibitem[\protect\citeauthoryear{{Kere{\v s}}, {Katz}, {Fardal}, {Dav{\'e}}  \&
  {Weinberg}}{{Kere{\v s}} et~al.}{2009}]{keres09}
{Kere{\v s}} D.,  {Katz} N.,  {Fardal} M.,  {Dav{\'e}} R.,   {Weinberg} D.~H.,
  2009, \mn@doi [\mnras] {10.1111/j.1365-2966.2009.14541.x}, \href
  {http://adsabs.harvard.edu/abs/2009MNRAS.395..160K} {395, 160}

\bibitem[\protect\citeauthoryear{{Klypin}, {Kravtsov}, {Bullock}, {Primack}  \&
  {Klypin}}{{Klypin} et~al.}{2001}]{Klypin2001}
{Klypin} A.,  {Kravtsov} A.~V.,  {Bullock} J.,  {Primack} C.~C.,   {Klypin} J.,
   2001, \mn@doi [\apj] {10.1086/321400}, \href
  {https://arxiv.org/abs/astro-ph/0006343v2} {554, 903}

\bibitem[\protect\citeauthoryear{{Kravstov}}{{Kravstov}}{1999}]{Kravstov99_thesis}
{Kravstov} A.~V.,  1999, PhD thesis, New Mexico State University

\bibitem[\protect\citeauthoryear{Kravtsov}{Kravtsov}{2003}]{Kravtsov_2003}
Kravtsov A.~V.,  2003, \mn@doi [The Astrophysical Journal] {10.1086/376674},
  590, L1

\bibitem[\protect\citeauthoryear{{Kravtsov}, {Klypin}  \&
  {Khokhlov}}{{Kravtsov} et~al.}{1997}]{Kravstov97}
{Kravtsov} A.~V.,  {Klypin} A.~A.,   {Khokhlov} A.~M.,  1997, \mn@doi [\apjs]
  {10.1086/313015}, \href {http://adsabs.harvard.edu/abs/1997ApJS..111...73K}
  {111, 73}

\bibitem[\protect\citeauthoryear{{Krogager}}{{Krogager}}{2018}]{krogager18}
{Krogager} J.-K.,  2018, arXiv e-prints, \href
  {https://ui.adsabs.harvard.edu/abs/2018arXiv180301187K} {p. arXiv:1803.01187}

\bibitem[\protect\citeauthoryear{{Krumholz} \& {Tan}}{{Krumholz} \&
  {Tan}}{2007}]{KrumholzTan07}
{Krumholz} M.~R.,  {Tan} J.~C.,  2007, \mn@doi [\apj] {10.1086/509101}, \href
  {https://ui.adsabs.harvard.edu/abs/2007ApJ...654..304K} {654, 304}

\bibitem[\protect\citeauthoryear{{Krumholz} \& {Thompson}}{{Krumholz} \&
  {Thompson}}{2012}]{KrumholzThompson12}
{Krumholz} M.~R.,  {Thompson} T.~A.,  2012, \mn@doi [\apj]
  {10.1088/0004-637X/760/2/155}, \href
  {https://ui.adsabs.harvard.edu/abs/2012ApJ...760..155K} {760, 155}

\bibitem[\protect\citeauthoryear{{Lehner}, {O'Meara}, {Fox}, {Howk},
  {Prochaska}, {Burns}  \& {Armstrong}}{{Lehner} et~al.}{2014}]{lehner14}
{Lehner} N.,  {O'Meara} J.~M.,  {Fox} A.~J.,  {Howk} J.~C.,  {Prochaska} J.~X.,
   {Burns} V.,   {Armstrong} A.~A.,  2014, \mn@doi [\apj]
  {10.1088/0004-637X/788/2/119}, \href
  {http://adsabs.harvard.edu/abs/2014ApJ...788..119L} {788, 119}

\bibitem[\protect\citeauthoryear{{Lehner}, {Wotta}, {Howk}, {O'Meara},
  {Oppenheimer}  \& {Cooksey}}{{Lehner} et~al.}{2018}]{Lehner18}
{Lehner} N.,  {Wotta} C.~B.,  {Howk} J.~C.,  {O'Meara} J.~M.,  {Oppenheimer}
  B.~D.,   {Cooksey} K.~L.,  2018, \mn@doi [\apj] {10.3847/1538-4357/aadd03},
  \href {https://ui.adsabs.harvard.edu/abs/2018ApJ...866...33L} {866, 33}

\bibitem[\protect\citeauthoryear{{Lehner}, {Wotta}, {Howk}, {O'Meara},
  {Oppenheimer}  \& {Cooksey}}{{Lehner} et~al.}{2019}]{lehner19}
{Lehner} N.,  {Wotta} C.~B.,  {Howk} J.~C.,  {O'Meara} J.~M.,  {Oppenheimer}
  B.~D.,   {Cooksey} K.~L.,  2019, \mn@doi [\apj] {10.3847/1538-4357/ab41fd},
  \href {https://ui.adsabs.harvard.edu/abs/2019ApJ...887....5L} {887, 5}

\bibitem[\protect\citeauthoryear{{Leitherer} et~al.,}{{Leitherer}
  et~al.}{1999}]{Leitherer99}
{Leitherer} C.,  et~al., 1999, \mn@doi [\apjs] {10.1086/313233}, \href
  {https://ui.adsabs.harvard.edu/abs/1999ApJS..123....3L} {123, 3}

\bibitem[\protect\citeauthoryear{{Liang} \& {Kravtsov}}{{Liang} \&
  {Kravtsov}}{2017}]{liang17}
{Liang} C.,  {Kravtsov} A.,  2017, arXiv e-prints, \href
  {https://ui.adsabs.harvard.edu/abs/2017arXiv171009852L} {p. arXiv:1710.09852}

\bibitem[\protect\citeauthoryear{{Liang}, {Kravtsov}  \& {Agertz}}{{Liang}
  et~al.}{2018}]{liang18}
{Liang} C.~J.,  {Kravtsov} A.~V.,   {Agertz} O.,  2018, \mn@doi [\mnras]
  {10.1093/mnras/sty1668}, \href
  {https://ui.adsabs.harvard.edu/abs/2018MNRAS.479.1822L} {479, 1822}

\bibitem[\protect\citeauthoryear{{Lodders}}{{Lodders}}{2019}]{Lodder19}
{Lodders} K.,  2019, arXiv e-prints, \href
  {https://ui.adsabs.harvard.edu/abs/2019arXiv191200844L} {p. arXiv:1912.00844}

\bibitem[\protect\citeauthoryear{{Lopez}, {Krumholz}, {Bolatto}, {Prochaska},
  {Ramirez-Ruiz}  \& {Castro}}{{Lopez} et~al.}{2014}]{Lopez14}
{Lopez} L.~A.,  {Krumholz} M.~R.,  {Bolatto} A.~D.,  {Prochaska} J.~X.,
  {Ramirez-Ruiz} E.,   {Castro} D.,  2014, \mn@doi [\apj]
  {10.1088/0004-637X/795/2/121}, \href
  {https://ui.adsabs.harvard.edu/abs/2014ApJ...795..121L} {795, 121}

\bibitem[\protect\citeauthoryear{{Lynch} \& {Charlton}}{{Lynch} \&
  {Charlton}}{2007}]{LynchCharlton07}
{Lynch} R.~S.,  {Charlton} J.~C.,  2007, \mn@doi [\apj] {10.1086/519826}, \href
  {https://ui.adsabs.harvard.edu/abs/2007ApJ...666...64L} {666, 64}

\bibitem[\protect\citeauthoryear{{Mar} \& {Bailey}}{{Mar} \&
  {Bailey}}{1995}]{mar95}
{Mar} D.~P.,  {Bailey} G.,  1995, \mn@doi [\pasa] {10.1017/S1323358000020324},
  \href {https://ui.adsabs.harvard.edu/abs/1995PASA...12..239M} {12, 239}

\bibitem[\protect\citeauthoryear{{Marra} et~al.,}{{Marra}
  et~al.}{2021a}]{marra21a}
{Marra} R.,  et~al., 2021a, \mn@doi [\mnras] {10.1093/mnras/stab2896}, \href
  {https://ui.adsabs.harvard.edu/abs/2021MNRAS.508.4938M} {508, 4938}

\bibitem[\protect\citeauthoryear{{Marra} et~al.,}{{Marra}
  et~al.}{2021b}]{marra21b}
{Marra} R.,  et~al., 2021b, \mn@doi [\apj] {10.3847/1538-4357/abd033}, \href
  {https://ui.adsabs.harvard.edu/abs/2021ApJ...907....8M} {907, 8}

\bibitem[\protect\citeauthoryear{{Masiero}, {Charlton}, {Ding}, {Churchill}  \&
  {Kacprzak}}{{Masiero} et~al.}{2005}]{masiero05}
{Masiero} J.~R.,  {Charlton} J.~C.,  {Ding} J.,  {Churchill} C.~W.,
  {Kacprzak} G.,  2005, \mn@doi [\apj] {10.1086/428426}, \href
  {https://ui.adsabs.harvard.edu/abs/2005ApJ...623...57M} {623, 57}

\bibitem[\protect\citeauthoryear{{Muzahid}, {Kacprzak}, {Churchill},
  {Charlton}, {Nielsen}, {Mathes}  \& {Trujillo-Gomez}}{{Muzahid}
  et~al.}{2015}]{muzahid15}
{Muzahid} S.,  {Kacprzak} G.~G.,  {Churchill} C.~W.,  {Charlton} J.~C.,
  {Nielsen} N.~M.,  {Mathes} N.~L.,   {Trujillo-Gomez} S.,  2015, \mn@doi
  [\apj] {10.1088/0004-637X/811/2/132}, \href
  {http://adsabs.harvard.edu/abs/2015ApJ...811..132M} {811, 132}

\bibitem[\protect\citeauthoryear{Nelson et~al.,}{Nelson
  et~al.}{2018}]{Nelson2018}
Nelson D.,  et~al., 2018, \mn@doi [Monthly Notices of the Royal Astronomical
  Society] {10.1093/mnras/sty656}, 477, 450

\bibitem[\protect\citeauthoryear{{Nelson} et~al.,}{{Nelson}
  et~al.}{2020}]{nelson20}
{Nelson} D.,  et~al., 2020, \mn@doi [\mnras] {10.1093/mnras/staa2419}, \href
  {https://ui.adsabs.harvard.edu/abs/2020MNRAS.498.2391N} {498, 2391}

\bibitem[\protect\citeauthoryear{{Nielsen}, {Kacprzak}, {Sameer}, {Murphy},
  {Nateghi}, {Charlton}  \& {Churchill}}{{Nielsen} et~al.}{2022}]{nielsen22}
{Nielsen} N.~M.,  {Kacprzak} G.~G.,  {Sameer} {Murphy} M.~T.,  {Nateghi} H.,
  {Charlton} J.~C.,   {Churchill} C.~W.,  2022, \mnras, submitted

\bibitem[\protect\citeauthoryear{{Oppenheimer}, {Dav{\'e}}, {Kere{\v s}},
  {Fardal}, {Katz}, {Kollmeier}  \& {Weinberg}}{{Oppenheimer}
  et~al.}{2010}]{oppenheimer10}
{Oppenheimer} B.~D.,  {Dav{\'e}} R.,  {Kere{\v s}} D.,  {Fardal} M.,  {Katz}
  N.,  {Kollmeier} J.~A.,   {Weinberg} D.~H.,  2010, \mn@doi [\mnras]
  {10.1111/j.1365-2966.2010.16872.x}, \href
  {http://adsabs.harvard.edu/abs/2010MNRAS.406.2325O} {406, 2325}

\bibitem[\protect\citeauthoryear{{Oppenheimer} et~al.,}{{Oppenheimer}
  et~al.}{2016}]{Oppenheimer16}
{Oppenheimer} B.~D.,  et~al., 2016, \mn@doi [\mnras] {10.1093/mnras/stw1066},
  \href {https://ui.adsabs.harvard.edu/abs/2016MNRAS.460.2157O} {460, 2157}

\bibitem[\protect\citeauthoryear{{Peeples} et~al.,}{{Peeples}
  et~al.}{2019}]{peeples19}
{Peeples} M.~S.,  et~al., 2019, \mn@doi [\apj] {10.3847/1538-4357/ab0654},
  \href {https://ui.adsabs.harvard.edu/abs/2019ApJ...873..129P} {873, 129}

\bibitem[\protect\citeauthoryear{{P{\'e}roux}, {Nelson}, {van de Voort},
  {Pillepich}, {Marinacci}, {Vogelsberger}  \& {Hernquist}}{{P{\'e}roux}
  et~al.}{2020}]{peroux20}
{P{\'e}roux} C.,  {Nelson} D.,  {van de Voort} F.,  {Pillepich} A.,
  {Marinacci} F.,  {Vogelsberger} M.,   {Hernquist} L.,  2020, \mn@doi [\mnras]
  {10.1093/mnras/staa2888}, \href
  {https://ui.adsabs.harvard.edu/abs/2020MNRAS.tmp.2703P} {}

\bibitem[\protect\citeauthoryear{{Petitjean} \& {Bergeron}}{{Petitjean} \&
  {Bergeron}}{1990}]{pb90}
{Petitjean} P.,  {Bergeron} J.,  1990, \aap, \href
  {http://adsabs.harvard.edu/abs/1990A\%26A...231..309P} {231, 309}

\bibitem[\protect\citeauthoryear{{Petitjean} \& {Bergeron}}{{Petitjean} \&
  {Bergeron}}{1994}]{pb94}
{Petitjean} P.,  {Bergeron} J.,  1994, \aap, \href
  {https://ui.adsabs.harvard.edu/abs/1994A&A...283..759P} {283, 759}

\bibitem[\protect\citeauthoryear{Pointon, Kacprzak, Nielsen, Muzahid, Murphy,
  Churchill  \& Charlton}{Pointon et~al.}{2019}]{Pointon_2019}
Pointon S.~K.,  Kacprzak G.~G.,  Nielsen N.~M.,  Muzahid S.,  Murphy M.~T.,
  Churchill C.~W.,   Charlton J.~C.,  2019, \mn@doi [The Astrophysical Journal]
  {10.3847/1538-4357/ab3b0e}, 883, 78

\bibitem[\protect\citeauthoryear{{Prochaska}}{{Prochaska}}{2003}]{prochaska03}
{Prochaska} J.~X.,  2003, \mn@doi [\apj] {10.1086/344595}, \href
  {https://ui.adsabs.harvard.edu/abs/2003ApJ...582...49P} {582, 49}

\bibitem[\protect\citeauthoryear{{Prochaska} et~al.,}{{Prochaska}
  et~al.}{2017}]{prochaska17}
{Prochaska} J.~X.,  et~al., 2017, \mn@doi [\apj] {10.3847/1538-4357/aa6007},
  \href {https://ui.adsabs.harvard.edu/abs/2017ApJ...837..169P} {837, 169}

\bibitem[\protect\citeauthoryear{{Rigby}, {Charlton}  \& {Churchill}}{{Rigby}
  et~al.}{2002}]{weakII}
{Rigby} J.~R.,  {Charlton} J.~C.,   {Churchill} C.~W.,  2002, \mn@doi [\apj]
  {10.1086/324723}, \href {http://adsabs.harvard.edu/abs/2002ApJ...565..743R}
  {565, 743}

\bibitem[\protect\citeauthoryear{{Rosenwasser}, {Muzahid}, {Charlton},
  {Kacprzak}, {Wakker}  \& {Churchill}}{{Rosenwasser}
  et~al.}{2018}]{rosenwasser18}
{Rosenwasser} B.,  {Muzahid} S.,  {Charlton} J.~C.,  {Kacprzak} G.~G.,
  {Wakker} B.~P.,   {Churchill} C.~W.,  2018, \mn@doi [\mnras]
  {10.1093/mnras/sty211}, \href
  {https://ui.adsabs.harvard.edu/abs/2018MNRAS.476.2258R} {476, 2258}

\bibitem[\protect\citeauthoryear{{Sameer} et~al.,}{{Sameer}
  et~al.}{2021}]{Sameer21}
{Sameer} et~al., 2021, \mn@doi [\mnras] {10.1093/mnras/staa3754}, \href
  {https://ui.adsabs.harvard.edu/abs/2021MNRAS.501.2112S} {501, 2112}

\bibitem[\protect\citeauthoryear{{Sameer} et~al.,}{{Sameer}
  et~al.}{2022}]{Sameer22}
{Sameer} et~al., 2022, \mn@doi [\mnras] {10.1093/mnras/stac052}, \href
  {https://ui.adsabs.harvard.edu/abs/2022MNRAS.510.5796S} {510, 5796}

\bibitem[\protect\citeauthoryear{{Savage} \& {Sembach}}{{Savage} \&
  {Sembach}}{1991}]{savage91}
{Savage} B.~D.,  {Sembach} K.~R.,  1991, \mn@doi [\apj] {10.1086/170498}, \href
  {http://adsabs.harvard.edu/abs/1991ApJ...379..245S} {379, 245}

\bibitem[\protect\citeauthoryear{{Schneider} et~al.,}{{Schneider}
  et~al.}{1993}]{schneider93}
{Schneider} D.~P.,  et~al., 1993, \mn@doi [\apjs] {10.1086/191798}, \href
  {http://adsabs.harvard.edu/abs/1993ApJS...87...45S} {87, 45}

\bibitem[\protect\citeauthoryear{{Steidel}}{{Steidel}}{1990}]{steidel90}
{Steidel} C.~C.,  1990, \mn@doi [\apjs] {10.1086/191493}, \href
  {https://ui.adsabs.harvard.edu/abs/1990ApJS...74...37S} {74, 37}

\bibitem[\protect\citeauthoryear{{Stocke}, {Keeney}, {Danforth}, {Shull},
  {Froning}, {Green}, {Penton}  \& {Savage}}{{Stocke} et~al.}{2013}]{stocke13}
{Stocke} J.~T.,  {Keeney} B.~A.,  {Danforth} C.~W.,  {Shull} J.~M.,  {Froning}
  C.~S.,  {Green} J.~C.,  {Penton} S.~V.,   {Savage} B.~D.,  2013, \mn@doi
  [\apj] {10.1088/0004-637X/763/2/148}, \href
  {http://adsabs.harvard.edu/abs/2013ApJ...763..148S} {763, 148}

\bibitem[\protect\citeauthoryear{{Strawn} et~al.,}{{Strawn}
  et~al.}{2021}]{strawn21}
{Strawn} C.,  et~al., 2021, \mn@doi [\mnras] {10.1093/mnras/staa3972}, \href
  {https://ui.adsabs.harvard.edu/abs/2021MNRAS.501.4948S} {501, 4948}

\bibitem[\protect\citeauthoryear{{Sutherland} \& {Dopita}}{{Sutherland} \&
  {Dopita}}{1993}]{sutherland93}
{Sutherland} R.~S.,  {Dopita} M.~A.,  1993, \mn@doi [\apjs] {10.1086/191823},
  \href {https://ui.adsabs.harvard.edu/abs/1993ApJS...88..253S} {88, 253}

\bibitem[\protect\citeauthoryear{{Tripp}, {Sembach}, {Bowen}, {Savage},
  {Jenkins}, {Lehner}  \& {Richter}}{{Tripp} et~al.}{2008}]{tripp08}
{Tripp} T.~M.,  {Sembach} K.~R.,  {Bowen} D.~V.,  {Savage} B.~D.,  {Jenkins}
  E.~B.,  {Lehner} N.,   {Richter} P.,  2008, \mn@doi [\apjs] {10.1086/587486},
  \href {https://ui.adsabs.harvard.edu/abs/2008ApJS..177...39T} {177, 39}

\bibitem[\protect\citeauthoryear{Trujillo-Gomez, Klypin, Colin, Ceverino,
  Arraki  \& Primack}{Trujillo-Gomez et~al.}{2015}]{Trujillo15}
Trujillo-Gomez S.,  Klypin A.,  Colin P.,  Ceverino D.,  Arraki K.~S.,
  Primack J.,  2015, \mn@doi [Monthly Notices of the Royal Astronomical
  Society] {10.1093/mnras/stu2037}, 446, 1140

\bibitem[\protect\citeauthoryear{{Vander Vliet}}{{Vander
  Vliet}}{2017}]{rachel_thesis}
{Vander Vliet} J.~R.,  2017, PhD thesis, New Mexico State University

\bibitem[\protect\citeauthoryear{{Welty}, {Hobbs}  \& {York}}{{Welty}
  et~al.}{1991}]{welty91}
{Welty} D.~E.,  {Hobbs} L.~M.,   {York} D.~G.,  1991, \mn@doi [\apjs]
  {10.1086/191537}, \href
  {https://ui.adsabs.harvard.edu/abs/1991ApJS...75..425W} {75, 425}

\bibitem[\protect\citeauthoryear{{Werk}, {Prochaska}, {Thom}, {Tumlinson},
  {Tripp}, {O'Meara}  \& {Peeples}}{{Werk} et~al.}{2013}]{werk13}
{Werk} J.~K.,  {Prochaska} J.~X.,  {Thom} C.,  {Tumlinson} J.,  {Tripp} T.~M.,
  {O'Meara} J.~M.,   {Peeples} M.~S.,  2013, \mn@doi [\apjs]
  {10.1088/0067-0049/204/2/17}, \href
  {http://adsabs.harvard.edu/abs/2013ApJS..204...17W} {204, 17}

\bibitem[\protect\citeauthoryear{{Werk} et~al.,}{{Werk} et~al.}{2014}]{werk14}
{Werk} J.~K.,  et~al., 2014, \mn@doi [\apj] {10.1088/0004-637X/792/1/8}, \href
  {http://adsabs.harvard.edu/abs/2014ApJ...792....8W} {792, 8}

\bibitem[\protect\citeauthoryear{{Wotta}, {Lehner}, {Howk}, {O'Meara}  \&
  {Prochaska}}{{Wotta} et~al.}{2016}]{Wotta16}
{Wotta} C.~B.,  {Lehner} N.,  {Howk} J.~C.,  {O'Meara} J.~M.,   {Prochaska}
  J.~X.,  2016, \mn@doi [\apj] {10.3847/0004-637X/831/1/95}, \href
  {https://ui.adsabs.harvard.edu/abs/2016ApJ...831...95W} {831, 95}

\bibitem[\protect\citeauthoryear{{Wotta}, {Lehner}, {Howk}, {O'Meara},
  {Oppenheimer}  \& {Cooksey}}{{Wotta} et~al.}{2019}]{Wotta19}
{Wotta} C.~B.,  {Lehner} N.,  {Howk} J.~C.,  {O'Meara} J.~M.,  {Oppenheimer}
  B.~D.,   {Cooksey} K.~L.,  2019, \mn@doi [\apj] {10.3847/1538-4357/aafb74},
  \href {https://ui.adsabs.harvard.edu/abs/2019ApJ...872...81W} {872, 81}

\bibitem[\protect\citeauthoryear{{Zahedy}, {Chen}, {Johnson}, {Pierce},
  {Rauch}, {Huang}, {Weiner}  \& {Gauthier}}{{Zahedy} et~al.}{2019}]{Zahedy19}
{Zahedy} F.~S.,  {Chen} H.-W.,  {Johnson} S.~D.,  {Pierce} R.~M.,  {Rauch} M.,
  {Huang} Y.-H.,  {Weiner} B.~J.,   {Gauthier} J.-R.,  2019, \mn@doi [\mnras]
  {10.1093/mnras/sty3482}, \href
  {https://ui.adsabs.harvard.edu/abs/2019MNRAS.484.2257Z} {484, 2257}

\bibitem[\protect\citeauthoryear{{Zahedy} et~al.,}{{Zahedy}
  et~al.}{2021}]{Zahedy21}
{Zahedy} F.~S.,  et~al., 2021, \mn@doi [\mnras] {10.1093/mnras/stab1661}, \href
  {https://ui.adsabs.harvard.edu/abs/2021MNRAS.506..877Z} {506, 877}

\bibitem[\protect\citeauthoryear{{Zonak}, {Charlton}, {Ding}  \&
  {Churchill}}{{Zonak} et~al.}{2004}]{zonak04}
{Zonak} S.~G.,  {Charlton} J.~C.,  {Ding} J.,   {Churchill} C.~W.,  2004,
  \mn@doi [\apj] {10.1086/382939}, \href
  {https://ui.adsabs.harvard.edu/abs/2004ApJ...606..196Z} {606, 196}

\bibitem[\protect\citeauthoryear{{van de Voort}, {Springel}, {Mandelker}, {van
  den Bosch}  \& {Pakmor}}{{van de Voort} et~al.}{2019}]{vandevoort19}
{van de Voort} F.,  {Springel} V.,  {Mandelker} N.,  {van den Bosch} F.~C.,
  {Pakmor} R.,  2019, \mn@doi [\mnras] {10.1093/mnrasl/sly190}, \href
  {https://ui.adsabs.harvard.edu/abs/2019MNRAS.482L..85V} {482, L85}

\makeatother
\end{thebibliography}


\bsp	
\label{lastpage}
\end{document}